%% file: 0_Main.tex
\documentclass[10pt,conference]{IEEEtran}

\usepackage[table]{xcolor}
\usepackage{balance}
\usepackage[utf8]{inputenc}
\usepackage[framemethod=TikZ]{mdframed}
\usepackage{tcolorbox}
\usepackage{graphicx}
\usepackage{float}
\usepackage{array}
\usepackage{tabularx}
\usepackage{multirow}
\usepackage{amsmath}
\usepackage{amssymb}
\usepackage{threeparttable}
\usepackage{setspace}
\usepackage{color}
\usepackage{colortbl}
\usepackage{xcolor}
\usepackage{soul}
\usepackage{booktabs}
\usepackage{makecell}
\usepackage{algorithm} 
\usepackage{algorithmic}
\usepackage{caption}
\usepackage{enumitem}
\usepackage{listings}
\usepackage{verbatim}
\usepackage{tikz}
\usepackage{picins}
\usepackage{arydshln}
\usepackage{subcaption}
\usepackage{parselines} 
\usepackage{wrapfig}
\usepackage{url}
\usepackage{cite}
\usepackage{pifont}
\usepackage{etoolbox, caption}
\usepackage[bottom]{footmisc}
\usepackage{hyperref}

\definecolor{hl-cream}{HTML}{FFF7CC}

\sethlcolor{hl-cream}
\soulregister\ref7
\soulregister\cite7
\soulregister\tool7 
\soulregister\pageref7
\soulregister\footnote7
\soulregister\eqref7

\renewcommand{\arraystretch}{1.2}

\usepackage[skip=1pt,labelfont=bf]{caption}
\usepackage{framed} 

\newcommand{\tool}{JUnitGenie}

\definecolor{customblue}{HTML}{006ca6}
\definecolor{customgreen}{HTML}{009264}
\definecolor{custombrown}{HTML}{ff3d00}
\AtEndPreamble{
 \usepackage{hyperref}
 \hypersetup{
  colorlinks = true,
  linkcolor = customblue,
  anchorcolor = customblue,
  citecolor = customgreen,
  filecolor = customblue,
  urlcolor = customblue
 }
}

\AtBeginDocument{%
  \providecommand\BibTeX{{%
    \normalfont B\kern-0.5em{\scshape i\kern-0.25em b}\kern-0.8em\TeX}}}

\newcommand{\find}[1]{
\begin{tcolorbox}[leftrule=1mm,toprule=0mm,bottomrule=0mm,left=1pt,right=2pt,top=2pt,bottom=2pt
]
\em #1
\end{tcolorbox}
}

\def\BibTeX{{\rm B\kern-.05em{\sc i\kern-.025em b}\kern-.08em
    T\kern-.1667em\lower.7ex\hbox{E}\kern-.125emX}}

\begin{document}

\title{Navigating the Labyrinth: Path-Sensitive Unit Test Generation with Large Language Models}

\author{
Dianshu Liao$^{1}$
\quad 
Xin Yin$^{2}$
\quad 
Shidong Pan$^{3}$
\quad 
Chao Ni$^{2}$
\quad 
Zhenchang Xing$^{3}$
\quad 
Xiaoyu Sun$^{1}$\IEEEauthorrefmark{1} 
\quad 
\\
\IEEEauthorblockA{
\textit{$^{1}$The Australian National University}, \{dianshu.liao, xiaoyu.sun1\}@anu.edu.au}
\IEEEauthorblockA{
\textit{$^{2}$The State Key Laboratory of Blockchain and Data Security, Zhejiang University}, \{xyin, chaoni\}@zju.edu.cn}
\IEEEauthorblockA{
\textit{$^{3}$CSIRO's Data61}, \{shidong.pan, zhenchang.xing\}@data61.csiro.au}
}

\maketitle

\begingroup
\renewcommand\thefootnote{\IEEEauthorrefmark{1}}
\footnotetext{Xiaoyu Sun is the corresponding author.}
\endgroup

\begin{abstract}

Unit testing is essential for software quality assurance, yet writing and maintaining tests remains time-consuming and error-prone. 
To address this challenge, researchers have proposed various techniques for automating unit test generation, including traditional heuristic-based methods and more recent approaches that leverage large language models (LLMs).
However, these existing approaches are inherently path-insensitive because they rely on fixed heuristics or limited contextual information and fail to reason about deep control-flow structures. As a result, they often struggle to achieve adequate coverage, particularly for deep or complex execution paths. In this work, we present a path-sensitive framework, \tool{}, to fill this gap by combining code knowledge with the semantic capabilities of LLMs in guiding context-aware unit test generation.
After extracting code knowledge from Java projects, \tool{} distills this knowledge into structured prompts to guide the generation of high-coverage unit tests. We evaluate \tool{} on 2,258 complex focal methods from ten real-world Java projects.
The results show that \tool{} generates valid tests and improves branch and line coverage by 29.60\% and 31.00\% on average over both heuristic and LLM-based baselines.
We further demonstrate that the generated test cases can uncover real-world bugs, which were later confirmed and fixed by developers.

\end{abstract}


\input{1_Introduction}

\input{2_Motivation}

\input{3_Approach}

\input{4_Evaluation}
\input{5_Discussion}
\input{6_Related_Work}

\input{7_Conclusion}

\bibliographystyle{unsrt}
\bibliography{sample-base}

\end{document}

%% file: 1_Introduction.tex
\section{Introduction}

Java remains a dominant programming language in 2025, with over 90\% of Fortune 500 companies employing it for their software development needs~\cite{netguru2025}. Ensuring code quality~\cite{sun2023lazycow}, identifying bugs~\cite{zhao2022autopatch,hu2023detecting,zhang2025fixing}, and maintaining code~\cite{sun2023taming} become challenging for Java developers as software grows in complexity. In practice, developers use unit testing (i.e., a white-box testing technique) as a cornerstone to ensure that their code is correctly implemented. Unit testing plays an important role in software development as: 1) it allows for early bug detection, avoiding exponential costs in production; 2) it is fast, meaning developers can run it frequently whenever a change is made without slowing down the development process; 3) it serves as a form of documentation, as every unit test explains how individual units of code are expected to behave.

Given the aforementioned advantages, ideally, every function in a Java software project should be covered by unit tests.
However, writing and maintaining unit tests is a non-trivial and time-consuming task, especially as modern software evolves rapidly~\cite{Daka2014ASO, Klammer2015WritingUT}. 
For many developers, it can take up around 15.8\% of total development time, adding a noticeable overhead to the development process~\cite{daka2014survey}. 
To tackle this challenge, researchers began exploring ways to automate the generation of unit tests, aiming to boost code coverage and improve development efficiency~\cite{fraser2011evosuite,DelgadoPrez2023InterEvoTRIE,Harman2010ATA, Csallner2008DySy,Ernst2007TheDS,Xiao2013CharacteristicSO,Pacheco2007FeedbackDirectedRT,fuzzingbook}.
For example, Fraser et al.~\cite{fraser2011evosuite} proposed EvoSuite, which uses evolutionary algorithms to generate test suites that maximize code and statement coverage. 
Although EvoSuite speeds up automation in unit test generation, it achieves only 28\% average branch coverage on real-world projects~\cite{Herlim2021EmpiricalSO}, struggling with methods that have semantic complexity (e.g., loops, recursion, non-trivial constraints). 
To address this problem, Xiao et al.~\cite{Xiao2013CharacteristicSO} further present an approach to leverage symbolic execution in enabling structural test generation.
Although the technique has been demonstrated effective on small to medium benchmarks, just like other symbolic execution tools, its scalability to large industrial codebases with deeply nested structure is limited.
In addition, with the rise of recent large language models (LLMs), researchers started to explore LLMs' capabilities in code understanding to automatically generate unit tests.
Specifically, Yuan et al.~\cite{Yuan2024EvaluatingAI} and Chen et al.~\cite{Chen2023ChatUniTestAF} simply feed LLMs with the implementation of focal methods (i.e., target methods for testing), together with the peer methods declared in the same class.
Their experiments demonstrate that LLMs are able to handle diverse programming structures as they are trained on a wide range of code examples.

To the best of our knowledge, the state-of-the-art (SOTA) LLM-based tools are only as good as the data they are trained on and do not possess true symbolic reasoning abilities, i.e., the ability to reason about program constraints such as determining input values to satisfy a branch condition.
As a result, they fail to generate tests that require precise reasoning about the program's behavior, leading to incorrect or incomplete test case generation. In addition, these approaches do not incorporate path-sensitive analysis, making it difficult to reason about multiple execution paths, especially when control-flow branches or deeply nested structures are involved. Moreover, these SOTA methodologies are often prompted with isolated method implementations with limited context, which limits their ability to generate meaningful unit tests and reach deeper execution paths.
Even with full context, the SOTA tools often require excessive prompt content, which distracts the models' focus from the focal method, resulting in poor code coverage and low-quality test generation~\cite{Liao2023A3CodGenAR, Shi2023LargeLM}. Indeed, despite these existing works, a fundamental question remains unresolved: \textbf{How can we enable path-sensitive unit test generation through concise LLM prompts?}

To address this, we design and implement \tool{}, a path-sensitive framework that guides LLMs in generating high-coverage test cases with minimal cognitive overhead, i.e., by supplying only the essential context needed while filtering out irrelevant information.
Specifically, we first build a code knowledge base (CKB) by extracting contextual knowledge (i.e., type information, control flow, and data dependencies) from a Java project under testing. 
We then traverse each method one control-flow path at a time, retrieve path-specific context from the CKB, and distill it into explicit instructions to guide test generation for that specific path. 
\tool{} applies multi-round refinement to resolve compilation and runtime errors, iteratively generating test cases that are both syntactically correct and executable.

Experimental results on thousands of methods under testing from real-world projects show that \tool{} is effective in generating test cases for Java projects.
It achieves a 69.88\% success rate in generating valid cases, outperforming both heuristic approaches~\cite{Fraser2011EvoSuiteAT, Pacheco2007RandoopFR} and LLM baselines~\cite{Yuan2024EvaluatingAI, Pizzorno2024CoverUpEH, Wang2024HITSHL}, with 16.02–48.00\% higher branch coverage and 20.08–49.25\% higher line coverage, and the most tests generated.
Our ablation study further reveals that simply providing abundant context can be counterproductive, as it may overwhelm the model and dilute its focus.
Combining contextual knowledge distillation with refinement achieves the best performance, yielding an additional 22.42\% improvement in branch coverage and 22.75\% in line coverage, highlighting the importance of path sensitivity and the need for context that is both sufficient and precise to generate executable tests. 
Moreover, we assess the generalizability of \tool{} across four foundation models. \tool{} consistently improves both branch and line coverage across all models by effectively leveraging structured context and iterative refinement.
It's also worth noting that \tool{} also uncovers four major bugs in real-world, which were promptly fixed by developers, demonstrating \tool{}'s practical usage in software development.

Overall, this work makes the following main contributions:
\begin{itemize}[leftmargin=*]
    \item We enable path-sensitive unit test generation by identifying three types of contextual knowledge: type information, control flow structure, and data dependencies, which can effectively guide LLMs.
    \item We present \tool{}, a reusable framework that extracts contextual code knowledge and distills it into explicit instructions to guide LLMs in generating high-coverage unit tests with minimal cognitive overhead.
    \item We conduct an extensive evaluation to demonstrate the effectiveness of \tool{}:
    (i) it significantly outperforms state-of-the-art tools across real-world Java projects;  (ii) ablation studies confirm the contribution of knowledge distillation and test case refinement;  (iii) the approach generalizes well across popular foundation models;  (iv) it successfully uncovers real bugs in production code.
\end{itemize}

The source code and experimental results are all made publicly available in our artifact package: \href{https://github.com/Dianshu-Liao/JUnitGenie}{[Link]}.

%% file: 2_Motivation.tex
\section{Motivation}\label{sec_motivation}

Our research fellows have put efforts in developing automatic unit test generation tools to enhance testing coverage, efficiency, ensuring overall software quality in Java applications~\cite{Fraser2011EvoSuiteAT, Pacheco2007RandoopFR, Yuan2024EvaluatingAI, Pizzorno2024CoverUpEH, Wang2024HITSHL}. These approaches
either leverage heuristic rules or AI-driven methods to achieve their purpose. Unfortunately, heuristic approaches rely on pre-defined rules to guide their test generation processes, lacking the capability of semantic reasoning, which makes them short of handling interdependent program constraints. For example, the SOTA search-based technique, EvoSuite~\cite{Fraser2011EvoSuiteAT} may explore large, irrelevant input spaces and stall in cases where precise values are needed, leading to poor coverage in such scenarios. Similarly, Randoop~\cite{Pacheco2007RandoopFR}  generates tests by randomly combining method calls without understanding the program's logic or the value of required inputs. This randomness makes it  unlikely to produce specific values or sequences needed to satisfy complex constraints. As a result, tests often fail early and provide limited coverage for deeper or more constrained code paths.

\begin{figure*}[!h]
    \centering
    \setlength{\abovecaptionskip}{6pt}
    \setlength{\belowcaptionskip}{-12pt}
    \includegraphics[width=0.78\textwidth]{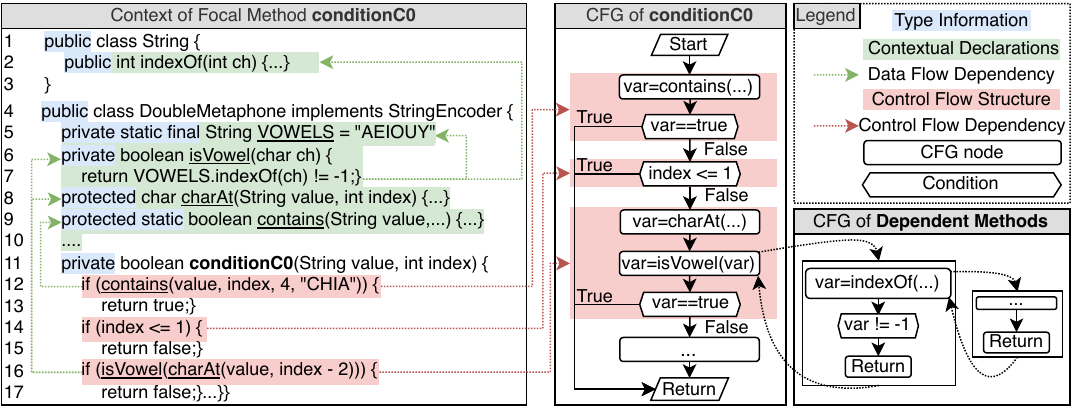}
    \caption{Motivating example.}
    \label{fig:motivating_example}
\end{figure*}

On the other hand, researchers have begun utilizing the semantic understanding capabilities of LLMs to generate test cases, as seen in representative works such as ChatTester~\cite{Yuan2024EvaluatingAI}, CoverUp~\cite{Pizzorno2024CoverUpEH}, and HITS~\cite{Wang2024HITSHL}.
While those LLM-based approaches show promising improvement when compared to traditional heuristic approaches, these approaches still struggle to achieve satisfactory code coverage on real-world projects~\cite{Gu2025LLMTG} for two main reasons: 
(1) they overlook critical information required to generate effective tests, such as type annotations, control-flow structures, and nested data-flow dependencies, forcing the model to infer missing semantics; and 
(2) they indiscriminately include the full implementation of the focal method (i.e., the core method being tested), together with all dependent method implementations without any filtering or relevance analysis. This overwhelms the LLM with excessive and irrelevant code, ultimately hindering its ability to focus and generate effective unit tests, especially for exercising deeper execution paths.

Therefore, we argue that there is a need to provide path-sensitive and concise
contextual information to LLMs to enable more effective and accurate test generation. We now present a concrete example in Fig.~\ref{fig:motivating_example} to further elaborate. Fig.~\ref{fig:motivating_example}  showcases a simplified code snippet excerpted from the Java project \href{https://github.com/apache/commons-codec/blob/master/src/main/java/org/apache/commons/codec/language/DoubleMetaphone.java}{\emph{commons-codec}}, incorporating the focal method \emph{DoubleMetaphone.conditionC0()} along with its calling context and dependent code knowledge. The left section outlines required contextual knowledge needed to generate executable unit tests for the API under testing. Specifically, lines 11 to 17 show the implementation of the focal method \emph{conditionC0()}, which includes three conditional statements. Its control flow logic is further detailed in the right section of Fig.~\ref{fig:motivating_example}, beginning with a method call to \emph{contains()} (i.e., line 9 in Fig.~\ref{fig:motivating_example}). If true, the method returns immediately. If not, it proceeds to the second conditional check on the value of index (i.e., line 14 in Fig.~\ref{fig:motivating_example}). When \emph{index $\le$ 1}, another early return is triggered. Otherwise, the flow continues by computing a character via method \emph{charAt()} (i.e., line 8 in Fig.~\ref{fig:motivating_example}) and passing it to \emph{isVowel()} (i.e., line 6 in Fig.~\ref{fig:motivating_example}). The final condition again determines whether the method returns true or false.

To generate executable unit tests for the focal method \emph{conditionC0()} with high coverage, one needs to incorporate three critical elements of contextual code knowledge: \textbf{(1) Type information}, which determines how the test should invoke the focal method. This includes handling access modifiers (e.g., private, protected, abstract), instantiating the host class based on its declaration, and considering inheritance and polymorphism (e.g., abstract superclasses, overridden methods, and dynamic dispatch);
\textbf{(2) Control-flow information}, which captures the full spectrum of execution paths. Each path is defined by conditional statements and requires tailored input values to achieve full coverage;
\textbf{(3) Data-flow dependencies}, which must be analyzed to construct valid inputs that satisfy path constraints. This involves understanding how values propagate through constants, fields, and method calls that influence the behavior of conditional branches.

\begin{figure*}[!h]
    \centering
    \setlength{\abovecaptionskip}{0.2pt}
    \setlength{\belowcaptionskip}{-15pt}
    \includegraphics[width=\textwidth]{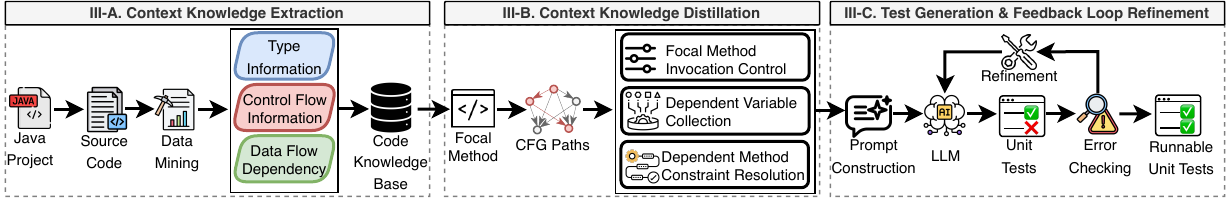}
    \caption{The working process of our approach.
    }
    \label{fig:approach_overview}
\end{figure*}

Given that traditional heuristic approaches (e.g., EvoSuite, Randoop) are designed on top of pre-defined rules or random strategies to explore input spaces, they struggle to reason about control flow or handle complex type constraints. 
For instance, EvoSuite fails to generate any valid tests for \emph{conditionC0()} as it cannot mutate inputs to satisfy the condition at line 12 in Fig.~\ref{fig:motivating_example}, causing the search algorithm to stall.
In addition, the SOTA LLM-driven methods (e.g., ChatTester, CoverUp and HITS), typically naively include the implementation of focal method with limited context in the prompt, without providing explicit instructions to guide the LLM in generating effective test cases. 
This distracts the model and leads it to ignore critical elements such as type information, control flow, and data dependencies, which are essential for achieving high coverage, resulting in invalid or shallow tests that fail to cover paths with complex constraints or dependencies.
For instance, \emph{CoverUp} fails to cover line 13 in Fig.~\ref{fig:motivating_example} due to missing control-flow and data-flow context.
Reaching this line requires inputs that bypass the condition on line 12, which depends on understanding the behavior of \emph{contains()} and requires explicit instruction in the prompt.
To summarize, the lack of deep understanding of the structural and semantic context of focal method hinders SOTA tools from achieving high coverage.
In this work, we propose to generate unit tests by learning contextual code knowledge and guiding LLMs with structured representations of type, control-flow, and data-flow information to enable effective and accurate test generation.

%% file: 3_Approach.tex
\section{Approach}~\label{sec_method}
The ultimate goal of our work is to enable path-sensitive unit test generation by providing minimal yet effective contextual code knowledge (i.e., type, control-flow, and data-flow information) to LLMs, thereby achieving better coverage. Fig.~\ref{fig:approach_overview} outlines the process of \tool{}, which is made up of three modules: \emph{Context Knowledge Extraction}, \emph{Context Knowledge Distillation}, and \emph{Test Generation with Feedback Loop Refinement}. 
We elaborate on each module below.

The first module, \emph{Context Knowledge Extraction}, takes as input a Java project and extracts comprehensive code knowledge from the source code.
Here, we collect \emph{Type information} (including method signature, access modifiers, class hierarchy information), and compile it into bytecode to build a call graph to retrieve control-flow information and data-flow dependencies. This extracted information is stored in a Code Knowledge Base (CKB), which serves as a foundation for subsequent test generation by enabling efficient, targeted context retrieval.

The second module, \emph{Context Knowledge Distillation}, generates executable calling contexts for all CFG paths of a focal method. 
Since a whole CFG may involve numerous constraints and dependent methods, exhaustively analyzing all paths can lead to exponential computational overhead and inflate the prompt which scatters the LLM's attention.
So \tool{} traverses each CFG path individually and retrieves the minimal required calling context (i.e., type information, control-flow structure, and data-flow dependencies) from the CKB. By isolating path-sensitive information, \tool{} reduces the prompt complexity and LLM reasoning burden. The distilled context enables path-sensitive reasoning by providing guidance that is both sufficient and concise for unit test generation.

The third module, \emph{Test Generation with Feedback Loop Refinement}, takes the distilled calling context as input and composes it into a structured prompt, which is then submitted to an LLM to generate unit test cases. As the generated tests may contain compilation or runtime errors (e.g., missing imports or incorrect class references), \tool{} captures error messages and iteratively refines the generated tests through a feedback loop until a fully executable test case is produced. We now
detail these three modules of \tool{} below.

\subsection{Context Knowledge Extraction} \label{sec_context_knowledge_extraction}
As shown in Fig.~\ref{fig:approach_overview}, the first module of \tool{}, namely \emph{Context Knowledge Extraction}, extracts structured code knowledge from a Java project to support subsequent modules. Specifically, it extracts the following information: 

1) Type information: Here, we leverage JavaParser~\cite{javaparser}, which can effectively extract both structural (e.g., method declarations) and semantic (e.g., inheritance relationships) code knowledge by parsing the source code. Specifically, we extract type information including method signatures (i.e., method names, return types, and parameter types), access modifiers (e.g., public, private), and class hierarchy information (e.g., superclass and implemented interfaces). These features are critical for understanding how caller instances can be properly instantiated or interacted with during test generation. For example, the access modifiers determine whether a method can be directly accessed or must be invoked reflectively. Similarly, class hierarchy information reveals whether a class is abstract (thus requiring a concrete subclass for instantiation) or whether a method is inherited or overridden, which affects how test inputs should be constructed and how method dispatch should be handled. This information ensures that generated tests respect the structural constraints of the code and can execute the focal methods in a valid runtime context.

2) Control-flow information: We then use SootUp~\cite{Karakaya2024SootUpAR}, which offers precise features for code analysis. In particular, it supports 3-address code intermediate representation Jimple and accurate call-graph construction algorithm Spark~\cite{lhotak2003scaling}. 
On top of SootUp, \tool{} decomposes Java programs into Jimple and decomposes expressions into control flow graph (CFG) paths, modeling method calls as atomic nodes and logical operators as branching edges. Each path has distinct guard conditions, enabling input synthesis for diverse execution scenarios. 
When extracting paths from a CFG, if a loop is encountered (i.e., a node reappears), we only keep the first iteration to preserve the loop’s logic, and skip subsequent visits within the same path to prevent infinite path extraction.

3) Data-flow dependencies: We again adopt SootUp to extract data-flow information such as variable definitions, field accesses, and dependent method calls to capture how values may influence branching conditions.
Specifically, dependent methods are those whose return values affect subsequent control decisions or outputs. 
For example, as shown in Fig.~\ref{fig:motivating_example}, the return value of \emph{contains()} (i.e., line 12) determines whether the method proceeds to the following \emph{return true} statement at line 13.
In addition, \tool{} also considers dependent variables, including parameters, local variables, and static fields whose values are referenced in conditional statements within the focal method. For instance, the static string variable \emph{VOWELS} is used in \emph{isVowel()} to determine whether a character belongs to the set of vowels, thereby influencing the outcome of the conditional branch at line 16. Understanding this context knowledge and their data flow enables LLM to assign appropriate values to variables when constructing test inputs. Together, these dependencies provide a detailed view of how data influences control flow, allowing \tool{} to generate inputs that satisfy path-specific conditions and fully exercise the focal method's behavior.

All structured code knowledge is then stored in a Neo4J graph database~\cite{Neo4J}, which supports efficient querying and retrieval by subsequent modules.
Rather than relying on ad-hoc analysis during test generation, the Code Knowledge Base (CKB) offers rapid access to relevant context, thereby improving both the scalability and accuracy of \tool{}.

\subsection{Context Knowledge Distillation} \label{sec_approach_context_knowledge_retrieval}
The second module statically constructs concise calling contexts for a given focal method, enabling LLMs to understand its execution behavior with minimal cognitive overhead. 
To achieve this, \tool{} leverages the control flow graph (CFG) to identify the contextual code knowledge required for the target method under test. Here, a CFG represents the control flows within a single method, where nodes correspond to statements (or a set of sequential statements), and edges represent control transitions such as conditionals and loops. The CFG enables path-sensitive analysis in \tool{} by explicitly capturing distinct execution branches.

\tool{} traverses each control-flow path of the focal method, capturing its diverse execution scenarios.
For each path, it selectively retrieves the relevant type, control-flow, and data-flow information specific to that path, ensuring the prompt is both sufficient and concise. Since a single path's contextual code knowledge (e.g., types, control flow, data dependencies) can be complex, triggering a specific statement (e.g., line 17 in Fig.~\ref{fig:motivating_example}) may require inputs that satisfy conditions defined earlier (e.g., line 16 in Fig.~\ref{fig:motivating_example}), which in turn depend on the return value of a dependent method (e.g., \emph{isVowel}). Including the full implementation of methods like \emph{isVowel} would overwhelm the prompt, distracting LLMs. To reduce complexity, we introduce a distillation step to prune the calling context, retaining only the minimal code required to trigger each target path. This is achieved through three strategies:

\paragraph{Focal Method Invocation Control}
Given that the JVM supports three types of method invocation (static call, special call, and virtual call), we design a targeted handling strategy to ensure correct invocation under different scenarios.
\begin{itemize} [leftmargin=*]
  \item \textbf{Static calls} are used for methods marked with the \emph{static} modifier.
  If the method is \emph{public}, it can be invoked directly via \emph{ClassName.method(...)}. For non-public static methods, reflective invocation is required to bypass access restrictions.
  \item \textbf{Special calls} cover constructors and private methods. Constructors are handled by creating instances via direct instantiation, while private methods are invoked via reflection~\cite{sun2021taming}.

  \item \textbf{Virtual calls} are instance method invocations that support polymorphism. 
  In such cases, the target method is resolved through dynamic dispatch based on 1) the runtime type of the receiver object and 2) the method signature at the call site. Here, the dynamic dispatch process traverses the class hierarchy of the receiver type to select the overriding method in the lowest subclass that matches the call site’s signature.
  If the resolved method is inaccessible from the test class (e.g., private, package-private, or protected), \tool{} marks the call for reflection to ensure correct invocation.

\end{itemize}

This strategy enables robust and accurate method invocation across diverse method types and access constraints, in accordance with JVM execution semantics.
Importantly, \tool{} targets methods at all access levels, as both public and non-public (e.g., private) methods may contain faults and thus require unit tests. Moreover, other SOTA heuristic \cite{Fraser2011EvoSuiteAT, Pacheco2007RandoopFR}  and LLM~\cite{Wang2024HITSHL, Pizzorno2024CoverUpEH, Yuan2024EvaluatingAI} baselines also support non-public methods via reflection\footnote{Our empirical study shows that, out of all generated tests, EvoSuite, Randoop, ChatTester, CoverUp, and HITS produced 338, 242, 915, 1,141, and 213 reflection-based tests for private focal methods, respectively.}, either explicitly (through configurations) or implicitly (through prompt design, where the LLM is instructed or trained to generate reflection code when needed).
Here, reflection is used to access methods not normally visible (e.g., private methods) so they can be invoked without modifying the source code. 
In our design, this stage relies entirely on static analysis. It examines method modifiers and class hierarchy to determine reflection needs. The resulting information is embedded into LLM prompts as guidance. When it comes to generating tests for private methods, the LLM inserts reflection code into generated tests, and these reflective calls are executed at test runtime.

\paragraph{Dependent Variable Collection}
Given a CFG path, its control flow may be influenced by variables defined outside the focal method. To ensure the intended execution of the path, it is necessary to assign appropriate values to these variable dependencies. \tool{} achieves this by identifying all referenced dependent variables along the path and collecting their declarations and access modifiers.  If a dependent variable is not \emph{public}, \tool{} instructs the model to use reflection to modify its value accordingly.

\paragraph{Dependent Method Constraint Resolution} \label{sec_dependent_method_constraint_resolution}
In the implementation of a focal method, the control flow may depend not only on variables but also on return values of other methods, referred to as dependent methods.
For path-sensitive test generation, it is crucial to construct inputs that can exercise specific execution paths within the focal method. 
To facilitate this, \tool{} first identifies the return statement of each dependent method and any intermediate statements along the call chain that may influence the returned value. 
\tool{} then traverses the CFG of each dependent method to identify all execution paths that are capable of producing the desired return value.
These paths are used to guide the generation of inputs capable of triggering the desired behaviors of the target path in the focal method.
As this step may still yield multiple feasible paths, \tool{} then goes a step deeper to select the simplest path, in order to minimize the LLM's reasoning time and resource usage.
Here, simplicity is defined by the following criteria: the selected path should involve the fewest additional dependent methods, rely on the minimal number of dependent variables, and have the fewest statements.
\begin{itemize}
    \item \textbf{Minimal Dependent Methods}: the selected path should involve the fewest additional dependent methods;
    \item \textbf{Minimal Dependent Variables}: the selected path should involve the minimal number of dependent variables;
    \item \textbf{Shortest Path Length}: the selected path should have the fewest statements.
\end{itemize}

This yields a control-flow path that satisfies the return constraints of the dependent method while minimizing the number of involved dependencies and the overall path complexity.

Given an optimal control-flow path within a dependent method, \tool{} determines parameter values necessary to follow that path and yield the expected return.
We formulate this task as a constraint-solving problem: based on the selected CFG path and return constraint, the LLM is prompted to infer parameter predicates that satisfy the path’s execution conditions.
If the selected path involves further dependent calls, \tool{} recursively applies the \emph{Dependent Method Constraint Resolution} process until all constraints are resolved.
For example, to reach the branch at line 16 in Fig.~\ref{fig:motivating_example}, \emph{isVowel(...)} must return \emph{true}. 
Since \emph{isVowel} returns \emph{true} when \emph{indexOf(...) $\ne -1$}, \tool{} selects a CFG path in \emph{indexOf} that ensures a return value $\ne -1$. 
It then derives the parameter constraints required to execute that path. 
Any input satisfying these constraints makes \emph{indexOf} return $\ne -1$, which in turn makes \emph{isVowel} return \emph{true}, thereby reaching the target branch.

When the same dependent method is invoked multiple times along a single CFG path, each invocation may require different return values for the CFG path to execute successfully. 
We analyze each invocation to determine what parameter constraints (e.g., parameter $\ge$ 3, parameter $<$ 5) would produce these required return values. 
Then we compute their intersection. If the intersection is empty, the path is discarded. If non-empty, the intersection (e.g., 3 $\le$ parameter $<$ 5) becomes the final parameter constraint, ensuring all invocations are satisfied.

\begin{figure*}
    \centering
    \setlength{\belowcaptionskip}{-10pt}
        \includegraphics[width=0.98\textwidth]{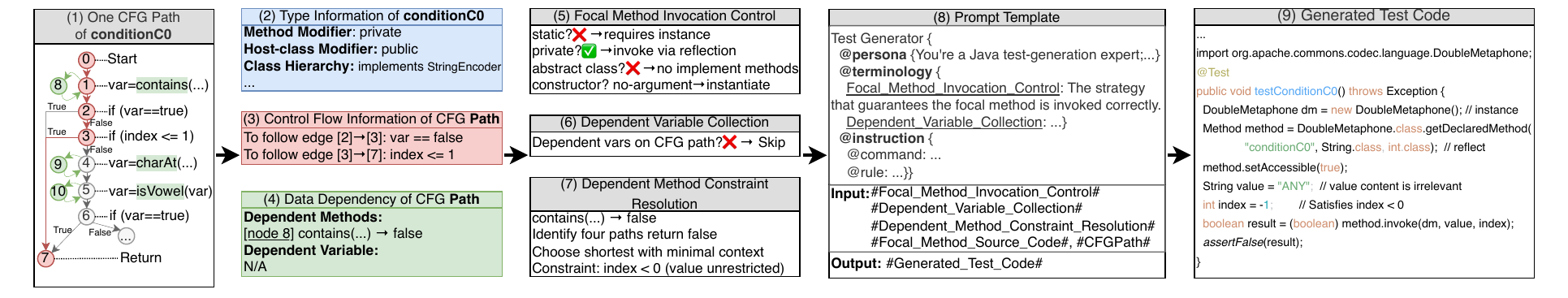}
    \caption{A running example of \tool{}.}
    \label{fig_running_example}
    \vspace{-2mm}
\end{figure*}

\subsection{Test Generation \& Feedback Loop Refinement} \label{sec_approach_test_gen_feedback_loop}
With the distilled context knowledge, \tool{} generates unit tests via an iterative pipeline where structured prompts are submitted to an LLM, and the outputs are refined through error checking until valid and runnable tests are obtained.

\paragraph{Prompt Construction}

The process begins with \emph{Prompt Construction}, where \tool{} transforms the distilled context knowledge into a structured prompt comprising three key components, as illustrated in Fig.~\ref{fig_running_example} (8):
\begin{itemize}[leftmargin=*]
  \item \textbf{@persona}: Instructs the model to behave as an expert Java developer tasked with writing unit tests.
  \item \textbf{@terminology}: Defines domain-specific concepts 
  to ensure consistent and precise references throughout the prompt.
  \item \textbf{@instruction}: Provides structured execution guidance:
  \begin{itemize}[leftmargin=*]
    \item \textbf{@command:} A four-step chain-of-thought (CoT) prompting strategy~\cite{Wei2022ChainOT}:
    (i) analyze the focal method and its CFG path;
    (ii) determine the correct invocation form (e.g., \emph{Class.method(...)} for static methods);
    (iii) create or modify variables to satisfy control-flow conditions; 
    (iv) construct input values that fulfill the parameter constraints of dependent methods.
    \item \textbf{@rule:} Enforces generation constraints, such as avoiding duplication of the focal method, assuming the test class resides in the \emph{test} directory, and using the JUnit library.
    \end{itemize}
\end{itemize}
To maintain input clarity and eliminate bias, \tool{} adopts a zero-shot prompting approach, which encourages the LLM to focus solely on the structured prompt without reference to in-context examples. 
Due to page limits, all AI prompt templates are provided in the \emph{prompts/} directory of the artifact~\cite{our_repo}.

\paragraph{Feedback Loop Refinement} \label{sec_feedback_loop_refinement}
The generated tests may fail to compile or trigger runtime exceptions, often due to inaccuracies in generation (e.g., hallucinated LLM-inferred parameter predicates, missing dependencies, or type mismatches). 
To address this, \tool{} applies a two-stage \emph{generate–validate–repair} loop with up to five refinement attempts per test (default).
1) \emph{Syntax validation and repair}: The generated test is first compiled using \emph{javac}~\cite{javac}.
If compilation fails, \tool{} collects the compiler diagnostics along with the faulty source code and forwards them to the \emph{Refinement} module, which produces a corrected version. This process repeats until the test compiles successfully.
2) \emph{Runtime validation and repair}:
Once compiled, the test is executed via JUnit to verify if the focal method is reachable. On runtime exceptions, the stack trace and source are sent to the \emph{Refinement} module for automated repair.
This process iterates until the test runs without error.
This step ensures that no exceptions are thrown before reaching the focal method, allowing it to be executed as intended.

\subsection{Running Example}
To illustrate \tool{}'s workflow, Fig.~\ref{fig_running_example} presents a step-by-step example for generating a test case targeting \emph{conditionC0}.
Fig.~\ref{fig_running_example} (1) shows its CFG, with the red-highlighted path [0]$\rightarrow$[1]$\rightarrow$[2]$\rightarrow$[3]$\rightarrow$[7].
To reduce LLM cognitive load, \tool{} processes one path at a time and retrieves three types of context: 
(2) type information, 
(3) control-flow constraints (e.g., to traverse edge [2]$\rightarrow$[3], the condition \emph{if (var == true)} must yield \emph{false}, hence \emph{var} must be \emph{false}); and (4) data dependencies (e.g., \emph{var} is the return value of \emph{contains(...)}, therefore \emph{contains(...)} must return \emph{false} when \emph{var} is required to be \emph{false}). These are distilled into structured instructions.
Since \emph{conditionC0} is non-static and private, it must be invoked via reflection as shown in (5).  
The host class is public, non-abstract, and has a no-argument constructor, so it can be instantiated with \emph{new DoubleMetaphone()}. 
Variable analysis in (6) finds no relevant data along this path and is skipped. 
\tool{} then resolves constraints for the dependent method \emph{contains()} in (7), which must return \emph{false} to satisfy the branch condition at node [2].
Among its six CFG paths, four return \emph{false}; \tool{} selects the shortest valid one, which returns \emph{false} when \emph{index} $<$ 0. 
Thus, \emph{index} must be negative, while \emph{value} is unconstrained.  
Finally, \tool{} assembles a structured prompt in (8) using the focal method, selected CFG path, and distilled context. 
Driven by this prompt, the LLM generates executable tests, one of which is shown in (9).

%% file: 4_Evaluation.tex
\section{Evaluation} \label{sec_evaluation}

\tool{} automates unit test generation by extracting and distilling contextual knowledge, enabling the LLM to reason about and explore deeper control- and data-dependent paths within the focal method. To assess its effectiveness, we propose to answer the following research questions (RQs):

\begin{itemize} [leftmargin=*] 
    \item \textbf{RQ1} How effective is \tool{} in generating executable unit test cases and how does it compare with existing tools?
    \item \textbf{RQ2} What types of contextual knowledge contribute most significantly to improving the performance of LLMs in automated test generation?
    \item \textbf{RQ3} How well does \tool{} generalize across different foundation models for automated test generation?
\end{itemize}

Experiments are conducted on a 16-core workstation with Intel Xeon Gold 6226R CPU @ 2.90GHz, 192GB RAM, and 8×NVIDIA A800 80GB GPUs, running Ubuntu 20.04.1 LTS.

\renewcommand{\arraystretch}{0.9}
\begin{table}[htbp]
\centering
\caption{\small Java projects used for evaluation.}
\label{tab_dataset}
\vspace{-0.5mm}
\resizebox{0.49\textwidth}{!}{%
\begin{tabular}{l|c|c|c|c}
\hline
\textbf{Project Name} & \textbf{Abbreviation} & \textbf{Version} & \textbf{\# Methods} & \textbf{\# Classes} \\
\hline
commons-codec         & CodeC       & \href{https://github.com/apache/commons-codec}{1.18.1}      & 132   & 45 \\
commons-collections   & Collections & \href{https://github.com/apache/commons-collections}{4.5.0-M4}    & 261   & 112 \\
commons-compress      & Compress    & \href{https://github.com/apache/commons-compress}{1.28.0}      & 294   & 133 \\
commons-csv           & CSV         & \href{https://github.com/apache/commons-csv}{1.13.1}      & 52    & 6 \\
jackson-core          & JCore       & \href{https://github.com/FasterXML/jackson-core}{2.19.0}      & 280   & 62 \\
jackson-databind      & JDataBind   & \href{https://github.com/FasterXML/jackson-databind}{2.19.0}      & 329   & 143 \\
jackson-dataformat-xml & JXML       & \href{https://github.com/FasterXML/jackson-dataformat-xml}{2.19.0}      & 113   & 22 \\
commons-jxpath        & JxPath      & \href{https://github.com/apache/commons-jxpath}{1.4}         & 247   & 75 \\
commons-lang          & Lang        & \href{https://github.com/apache/commons-lang}{3.18.0}      & 283   & 70 \\
joda-time             & JodaTime        & \href{https://github.com/JodaOrg/joda-time}{2.13.1}      & 267   & 94 \\
\hline
Total                 & -           & -           & 2,258 & 762 \\
\hline
\end{tabular}
}
\vspace{-0.5mm}
\end{table}
\renewcommand{\arraystretch}{1.0}

\subsection{Dataset} \label{sec_dataset}

To evaluate the effectiveness of \tool{}, we use real-world Java projects from Defects4J~\cite{Just2014Defects4JAD}, a widely adopted benchmark for automated testing research.
We use the latest versions of these projects to reflect modern development practices.
In addition, we observe Defects4J includes 17 Java projects with varying JDK requirements. To ensure a consistent and controlled experimental environment, we standardize on JDK 8 as the target platform, given its continued prevalence in open-source libraries and enterprise systems~\cite{Mak2024AutomaticBR, Watkinson2024ComparingAA, Schott2024JavaBN}. Based on this criterion, we exclude five projects (JFreeChart, Commons-CLI, Gson, Jsoup, and Math) that require JDK versions beyond 8 and cannot be compiled under our configuration. Additionally, we exclude two projects (Mockito and Closure-Compiler) that are not Maven-based and therefore cannot be built using our toolchain. As a result, 10 Java projects remain and are used as our experimental dataset, as shown in Table~\ref{tab_dataset}.

Within these 10 projects, not all methods are equally meaningful for evaluating test generation tools.
Specifically, methods with trivial logic (e.g., no branching or dependencies) can be easily covered by naive approaches and fail to showcase the strengths of advanced tools.
To ensure rigorous evaluation, we focus on methods that are both challenging and representative of real-world complexity.
Specifically, we select focal methods that meet two criteria:
1) contain at least one branching condition; and
2) have data dependencies on variables or methods defined outside its own body.
These filters yield 8,367 candidate methods across the 10 projects. 
However, generating tests for all would incur significant computational and token costs.
To balance cost with statistical significance, we randomly sample\footnote{The sample size is determined based on a confidence level at 95\% and a confidence interval at 5 (https://www.surveysystem.com/sscalc.htm).} a subset of methods from each project, as summarized in Table~\ref{tab_dataset}. Hence, the overall dataset contains 2,258 focal methods spread across 762 classes.

\subsection{{RQ1-Effectiveness}} \label{sec_evaluation_rq1}
Our first research question concerns the effectiveness of \tool{} in test case generation. To this end, we evaluate \tool{} by measuring its ability to produce valid tests and comparing its performance with existing SOTA tools. Here, we consider a test
case to be valid if (1) the generated code snippet can be successfully
compiled and (2) it does not throw an
exception before the execution point of the focal method. The first condition ensures that the test case is syntactically correct. The second condition guarantees that the execution context of the focal method under test is properly constructed so that the method can be successfully reached during execution.

In addition, for the comparison with SOTA tools, we include both heuristic test case generation approaches (i.e., EvoSuite~\cite{Fraser2011EvoSuiteAT} and Randoop~\cite{Pacheco2007RandoopFR}) and LLM-driven methods (i.e., ChatTester~\cite{Yuan2024EvaluatingAI}, CoverUp~\cite{Pizzorno2024CoverUpEH}, HITS~\cite{Wang2024HITSHL}) as baselines to evaluate our approach. 
We run EvoSuite v1.2.0 and Randoop v4.3.3 with default configurations, and execute all LLM-based methods using GPT-4o-Mini~\cite{GPT4omini} as the foundation model.

To compute coverage, we discard non-compilable tests and then measure coverage with \href{https://www.eclemma.org/jacoco/}{JaCoCo}, consistent with prior studies \cite{Yuan2024EvaluatingAI, Wang2024HITSHL}. We then break down the results as follows:

\begin{table}[htbp]
\centering
\caption{\small Number of valid test cases generated by each tool.}
\label{tab_valid_test_num_horizontal}
\vspace{-0.5mm}
\resizebox{0.49\textwidth}{!}{
\begin{tabular}{ccccccc}
\toprule
\textbf{Tool} 
& \textbf{EvoSuite} 
& \textbf{Randoop} 
& \textbf{ChatTester} 
& \textbf{CoverUp} 
& \textbf{HITS} 
& \textbf{\tool{}} \\
\midrule
\textbf{\makecell[l]{\# Generated Tests}} 
& 3,398  
& 4,192
& 6,324 
& 7,971 
& 8,351 
& \textbf{20,305} \\

\textbf{\makecell[l]{\# Valid Tests}} 
& 3,232 
& 4,018 
& 3,136
& 3,958 
& 4,756 
& \textbf{14,190} \\
\midrule

\textbf{\makecell[l]{\# Success Rate}} 
& 95.11\% 
& 95.85\% 
& 49.59\% 
& 49.65\%
& 56.95\%
& 69.88\% \\

\bottomrule
\end{tabular}
}

\end{table}

\begin{figure}[t]
    \centering
    \begin{subfigure}{0.241\textwidth}
        \centering
        \includegraphics[width=\textwidth]{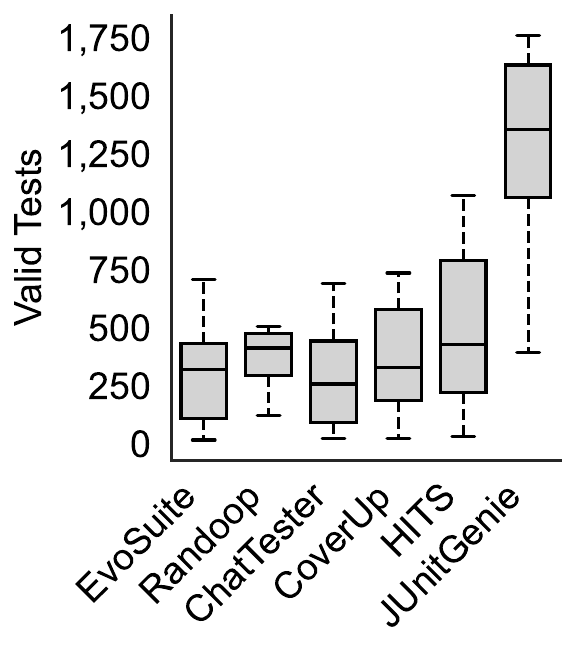}
        \vspace{-7mm}
        \caption{}
        \label{fig_valid_test_comparison}
    \end{subfigure}
    \hfill
    \begin{subfigure}{0.241\textwidth}
        \centering
        \includegraphics[width=\textwidth]{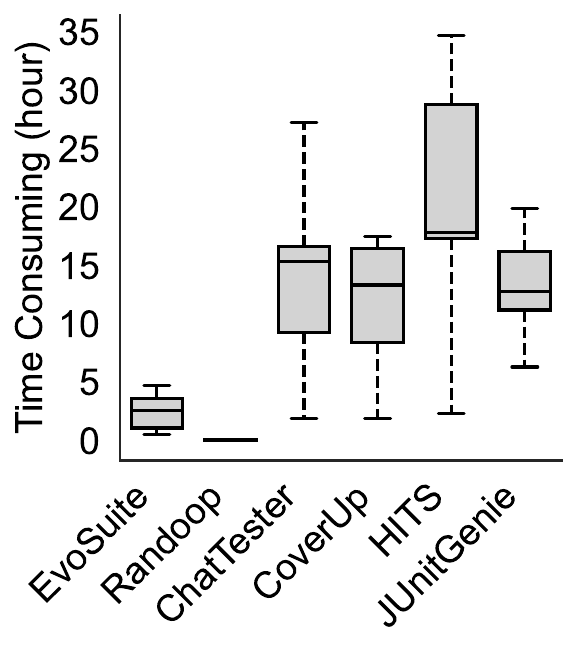}
        \vspace{-7mm}
        \caption{}
        \label{fig_gen_time}
    \end{subfigure}
    \caption{Comparison of (a) valid generated tests and (b) test generation time across all projects.}
    \label{fig_rq1_runtime_passrate}

\end{figure}

\input{Tables/Comparison_With_Baselines}

\textbf{Result.} 
Across 2,258 focal methods, \tool{} identifies and analyzes 14,132 CFG paths, with an average of 6.26 per method. 
From these paths, \tool{} generates 20,305 test cases.
After compilation and execution, 14,190 are identified as valid (i.e., 6,115 of them are invalid), giving a success rate of 69.88\% in generating valid test cases. Compared to other LLM-based tools, \tool{} achieves both the highest number of valid test cases and the highest success rate. While heuristic-based tools such as EvoSuite and Randoop attain higher success rates (95.11\% and 95.85\%, respectively), their total number of valid test cases they produce is significantly lower. This indicates that \tool{} strikes a better balance between generation volume and test validity, demonstrating its practical effectiveness in automated test generation at scale.
For fairness, we verified that the baseline tools also generate tests that explicitly treat private methods as focal methods and invoke them via reflection, consistent with \tool{}.

In addition, Table~\ref{tab_valid_test_num_horizontal} reports the number of valid test cases generated by each tool.
Specifically, EvoSuite and Randoop generate 3,232 and 4,018 valid tests, while LLM-based tools ChatTester, CoverUp, HITS, and \tool{} produce 3,136, 3,958, 4,756, and 14,190 valid tests, respectively.
These results highlight the superior effectiveness of \tool{} in generating executable test cases.

Fig.~\ref{fig_valid_test_comparison} further summarizes the distribution of valid test cases per project for each tool.
Overall, LLM-based tools tend to produce more valid test cases than heuristic-based approaches. 
Specifically, the median number of valid test cases per project is 323 for EvoSuite, 415 for Randoop, 260 for ChatTester, 332 for CoverUp, 432 for HITS, and 1,355 for \tool{}, with \tool{} also exhibiting the highest upper quartile at 1,633. 
Like most other SOTA approaches, \tool{} may fail on focal methods with complex control or data dependencies.
Nonetheless, \textbf{\tool{} is still capable of generating more valid test cases than existing tools}, demonstrating its effectiveness in leveraging distilled context across diverse codebases.

Moreover, we compare the branch and line coverage achieved by each tool across the 10 Java projects, as shown in Table~\ref{tab_comparison_with_baseline}. \tool{} consistently achieves the highest average coverage, with 56.86\% for branch and 61.45\% for line coverage, outperforming both heuristic and LLM-based baselines. It delivers particularly strong results on complex projects such as \emph{Lang} (87.10\% branch coverage, 92.41\% line coverage) and \emph{CSV} (76.57\% branch coverage, 85.98\% line coverage), demonstrating its ability to generate tests that exercise diverse and deep execution paths. Although tools like EvoSuite and Randoop perform well on a few individual projects, their overall average coverage remains substantially lower. Among LLM-based methods, HITS achieves the second-best performance with 32.13\% branch and 36.45\% line coverage, falling notably behind \tool{}. These results further validate the effectiveness of \tool{} in producing high-coverage test suites across a range of real-world Java applications. We then break down the comparative results as follows:

We further manually inspect the invalid test cases generated by existing tools and observe that their reliance on fixed heuristics and random input generation limits their ability to reason about control flow or handle complex type and context constraints. As a result, these tools often fail to explore deeper execution paths and tend to produce trivial or non-executable tests when the required input conditions are not satisfied. LLM-based methods, while flexible, often suffer from high rates of invalid test generation due to insufficient and undistilled context. They commonly omit key elements such as dependent variables or nested method calls, which forces the model to infer missing details and often results in incorrect reasoning. Also, they lack explicit guidance on aspects like using reflection for private methods or targeting specific branches, leading to poor reasoning and test failures. In contrast, \tool{} combines context-aware knowledge extraction and distillation with an iterative refinement loop that repairs invalid tests. This combination enables \tool{} to generate substantially more usable and executable test cases.

\begin{figure}[t]
    \centering
    \begin{subfigure}{0.241\textwidth}
        \centering
        \includegraphics[width=\textwidth]{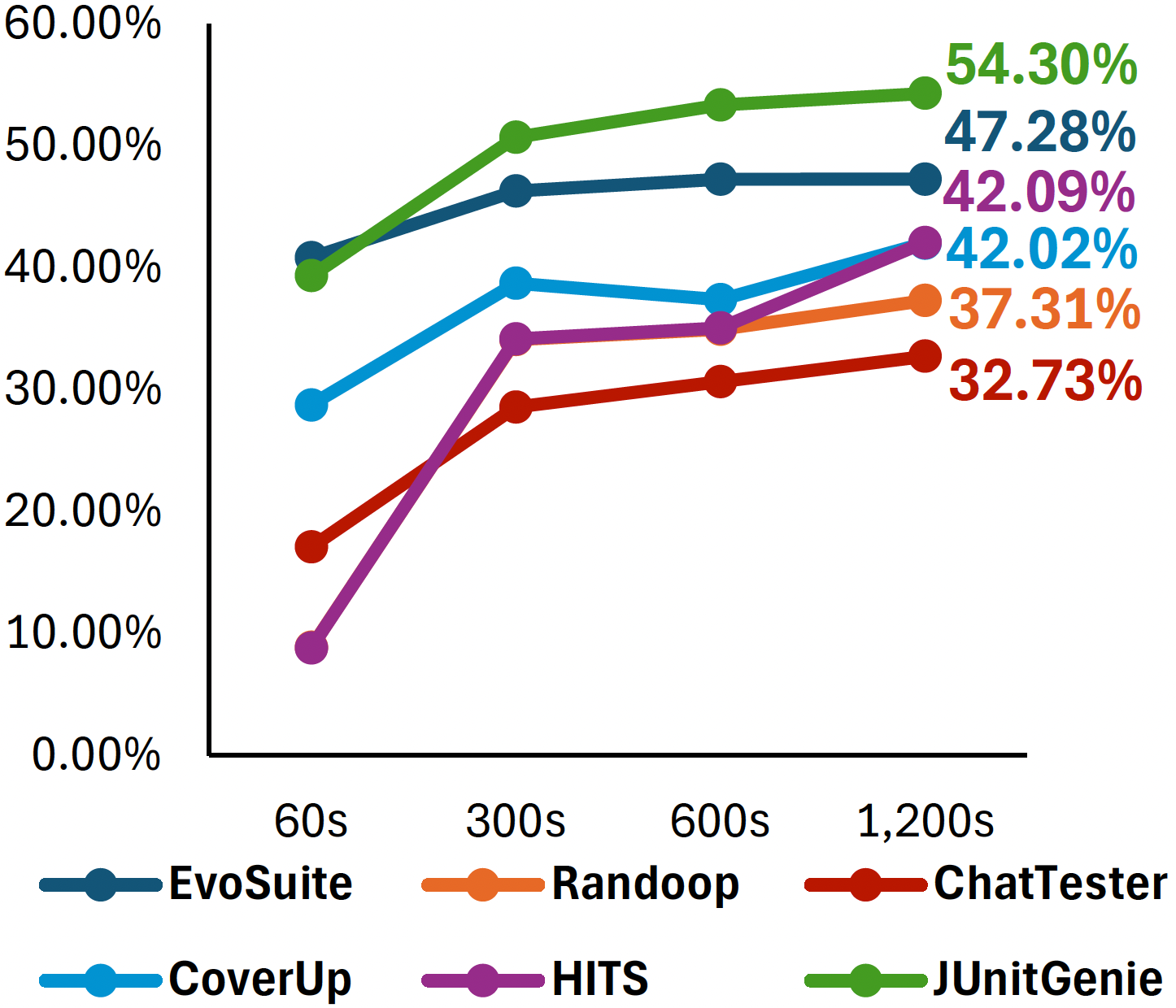}
        \vspace{-6mm}
        \caption{}
        \label{fig_avg_branch_coverage}
    \end{subfigure}
    \hfill
    \begin{subfigure}{0.241\textwidth}
        \centering
        \includegraphics[width=\textwidth]{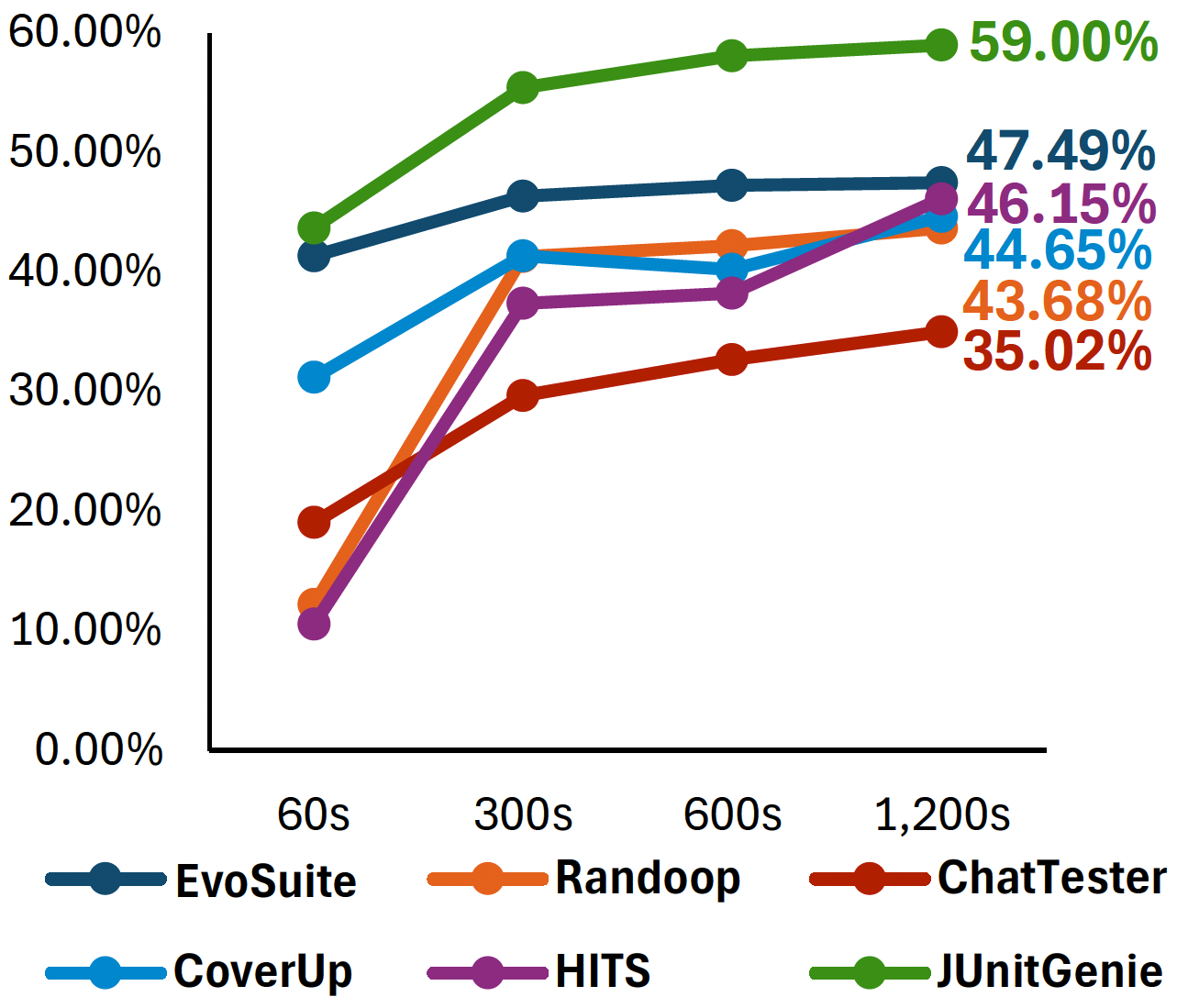}
        \vspace{-6mm}
        \caption{}
        \label{fig_avg_line_coverage}
    \end{subfigure}
    \caption{Comparison of the average (a) branch coverage and (b) line coverage under different time budgets across all projects.}
    \label{fig_discussion_study}
    \vspace{-1mm}
\end{figure}

\textbf{Comparison with baselines under unified time budgets.}
To ensure fairness, we further evaluate JUnitGenie and baselines under the same time budgets.
To this end, we run all tools on the entire dataset, applying timeouts of 60s, 300s, 600s, and 1,200s for each focal method.
Fig.~\ref{fig_discussion_study} presents the comparative results. 
Overall, JUnitGenie outperforms all baselines in branch and line coverage except at 60s, where EvoSuite is slightly better.
In addition, its advantage becomes more pronounced with longer budgets, indicating that JUnitGenie scales more effectively with extended testing time and achieves higher overall coverage across real-world projects.

With a 60s timeout, EvoSuite achieves slightly higher branch coverage (40.84\%) and comparable line coverage (41.37\%) to JUnitGenie  (39.33\% branch coverage, 43.72\% line coverage), while other tools perform noticeably lower. 
Here, EvoSuite achieves quick gains under short budgets through heuristic-driven evolutionary search, whereas JUnitGenie requires more time due to its path-sensitive CFG analysis. 
This situation is dramatically changed when the timeout is extended to 300s and 600s, where JUnitGenie begins to outperform all baselines on both branch and line coverage. 
Given more time, JUnitGenie resolves more complex path constraints and refines outputs through feedback loops that fix early syntax or runtime errors, improving effectiveness.
In contrast, heuristic tools (i.e., EvoSuite and Randoop) quickly reach their achievable coverage, and additional time provides little benefit for complex paths. 
For example, from 300s to 600s, EvoSuite’s branch and line coverage increased by only 0.69\% and 0.91\%, while Randoop’s increased by just 0.83\% and 0.87\%.
LLM-based baselines also improve with longer budgets (e.g., at 1,200s HITS reaches 42.09\% branch and 46.15\% line), but consistently underperform JUnitGenie due to lacking distilled context and explicit path guidance.
When the timeout is extended to 1,200s, the improvements of all tools gradually converge and JUnitGenie still achieves the highest coverage on both metrics (54.30\% branch coverage and 59.00\% line coverage), highlighting its better performance.

\textbf{Efficiency.} We further look at the efficiency of \tool{} by reporting its time consuming when generating test cases. 
As shown in Fig.~\ref{fig_gen_time}, Randoop is the fastest, followed by EvoSuite, which completes within 5 hours per project. All LLM-based methods require over 10 hours on average, with \tool{} taking around 13.68 hours per project. 
The main bottleneck lies in the code refinement phase, which accounts for over 60\% of the total time.
In terms of median generation time, ChatTester, CoverUp, and HITS take 15.40, 13.35, and 17.85 hours, respectively, while \tool{} requires 12.80 hours.
The fact that \tool{} spends relatively less time than other LLM-based tools suggests that it achieves a practical balance between efficiency and effectiveness, while producing the highest number of valid tests among all methods.

\find{\textbf{Answers to RQ1:}
\tool{} outperforms SOTA heuristic tools (e.g., EvoSuite) and other LLM-based approaches by generating more valid test cases and achieving higher line and branch coverage on real-world Java projects.
}

\input{Tables/Ablation_Study}

\begin{figure}
    \setlength{\abovecaptionskip}{8pt}
    \centering
    \includegraphics[width=0.49\textwidth]{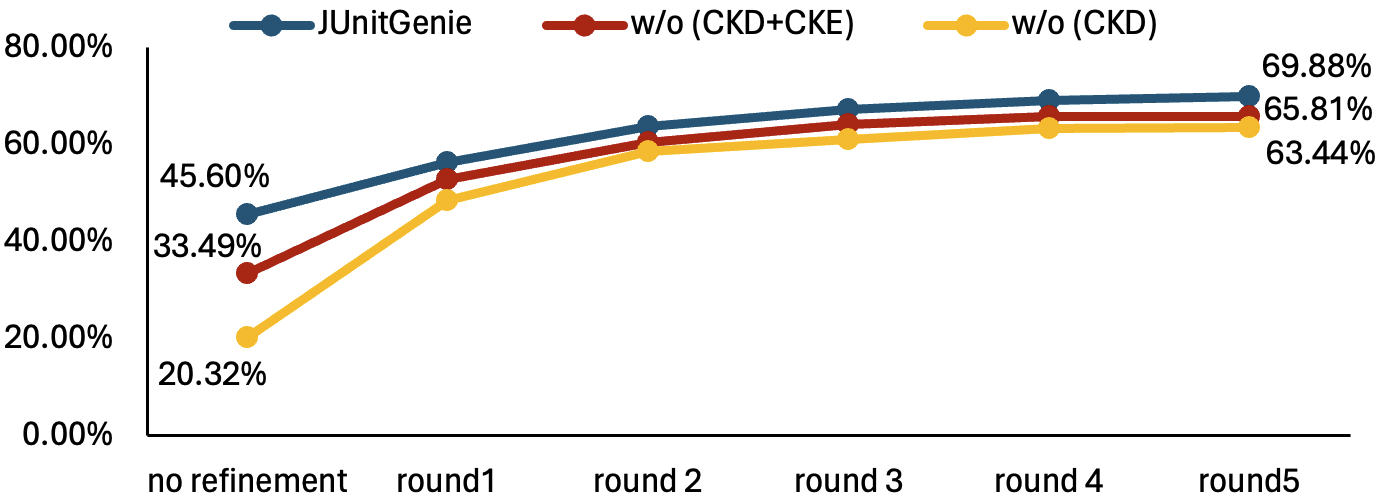}
    \caption{Evolution of valid test rate over refinement rounds.}
    
    \label{fig_runnable_tests_percentage}
\end{figure}

\subsection{RQ2-Knowledge Acquisition and Utilization} \label{sec_eva_ablation_study}
While \tool{} demonstrates strong performance in generating high-coverage test cases, the specific contributions of its core components remain unclear. To answer this question, we conduct an ablation study to assess the impact of three key steps: Context Knowledge Extraction (CKE), Context Knowledge Distillation (CKD), and Test Case Refinement (TCR). The study aims to clarify how each component contributes to the overall effectiveness of the test generation process.

Here, we design four ablation baselines to assess the contribution of each component in \tool{}. The most basic variant (\emph{w/o CKD+CKE+TCR}) removes all modules, providing only the focal method and its signature. The second (\emph{w/o CKD+CKE}) retains only test case refinement, using error messages to repair tests without any contextual input. The third (\emph{w/o CKD}) includes both context extraction and refinement but omits distillation, presenting raw context without structured guidance. The full \tool{} configuration includes all three components. All baselines use GPT-4o-Mini and are evaluated on 2,258 focal methods from 10 real-world Java projects.

\input{Tables/Different_Foundation_Models}

\textbf{Result.} 
Table~\ref{tab_ablation_study} presents the performance of four ablation baselines, each designed to evaluate the individual contribution of a specific component of \tool{}. The \emph{w/o (CKD+CKE+TCR)} setting performs the worst, achieving only 24.74\% branch coverage and 27.84\% line coverage. Without contextual knowledge and refinement support, the model operates with minimal information, relying solely on the focal method. This limitation often leads to invalid tests caused by unresolved variables, incorrect API usage, or missing imports.

Adding test case refinement (\emph{w/o CKD+CKE}) improves coverage by 12–15\%, as it repairs syntactically invalid yet semantically meaningful tests through iterative refinement of compiler error messages.
As illustrated by the red line in Fig.~\ref{fig_runnable_tests_percentage}, refinement raises the executable test rate from 33.49\% to 65.81\% over five iterations. 
Other variants show substantial gains over five rounds, with \emph{\tool{}} improving from 45.60\% to 69.88\%, and \emph{w/o (CKD)} from 20.32\% to 63.44\%. 
These trends highlight the critical role of iterative refinement in recovering test cases, especially when combined with contextual knowledge. It is also worth noting that \tool{}, which incorporates both contextual knowledge and refinement, achieves the highest improvement, reaching 69.88\% by round five.
In comparison, \emph{w/o (CKD+CKE)} and \emph{w/o (CKD)} plateau at 65.81\% and 63.44\%, respectively.
These results highlight the importance of both context and refinement: while refinement alone substantially boosts test executability, the combination with contextual knowledge yields the most effective and consistent improvements.

Notably, enabling context extraction without distillation (\emph{w/o CKD}) results in a slight performance drop, with branch coverage decreasing by 3.21\% and line coverage by 4.27\% compared to using refinement alone.
This decline is due to the overhead of excessive, unfiltered information.
For example, the method \href{https://github.com/JodaOrg/joda-time/blob/main/src/main/java/org/joda/time/chrono/LimitChronology.java}{\emph{withZone(DateTimeZone)}}
in the \emph{org.joda.time} project involves 36 execution paths, 40 dependent methods, and 20 variables, yielding 9,489 tokens of raw context. The excessive volume of undistilled context hinders the model’s ability to isolate and leverage relevant information, leading to test generation failure and zero coverage for the method.

Context knowledge distillation yields a significant improvement in performance. \tool{} attains 56.86\% branch coverage and 61.45\% line coverage, representing gains of 22.42\% and 22.75\% over \emph{w/o CKD}, respectively. These results highlight that simply providing abundant context is ineffective; the information must be distilled into a structured form to guide the generation process effectively. Effective integration of contextual knowledge depends not only on its completeness, but also on its accuracy and relevance.

\find{\textbf{Answers to RQ2:}
The combination of contextual knowledge distillation and refinement yields the best results, enabling path-sensitive test generation and showing the effectiveness of targeted guidance for producing executable tests.
}

\subsection{RQ3-Model Generality}

The previous two research questions evaluated \tool{} using GPT-4o-Mini, leaving it unclear whether the observed effectiveness stems from the method itself or the model. To address this, we investigate whether \tool{} consistently improves test generation across different foundation models.

We evaluate \tool{} across four foundation models: GPT-4o-Mini~\cite{GPT4omini}, and three widely adopted open-source code models: CodeLlama~\cite{Rozire2023CodeLO}, DeepSeek-Coder~\cite{Guo2024DeepSeekCoderWT}, and Qwen2.5-Coder~\cite{Hui2024Qwen25CoderTR}. Specifically, we use the 34B version of CodeLlama, the 33B version of DeepSeek-Coder, and the 32B version of Qwen2.5-Coder, all running in fp16 precision.

All models are configured with a maximum input length of 8,192 tokens and an output limit of 2,048 tokens.
For each model, we compare two settings: (1) Basic Prompting, which provides only the focal method's source code; and (2) \tool{}'s structured prompting, which applies the full pipeline in Section~\ref{sec_method} to build inputs enriched with distilled contextual knowledge.
The evaluation uses a shared benchmark of 2,258 focal methods from 10 Java projects.

\textbf{Result.}
We report branch and line coverage across four foundation models in Table~\ref{tab_different_foundation_model_reordered}.
The results show that our method consistently outperforms the basic setting on all models. For instance, using basic prompting, GPT-4o-Mini achieves only 24.74\% branch coverage, whereas \tool{} raises it to 56.86\%. 
Similar gains are observed across other models, confirming the broad effectiveness of our approach.

Among the three open-source models evaluated with our approach, Qwen2.5-Coder achieves the best overall results with 56.38\% branch and 60.68\% line coverage, followed by DeepSeek-Coder and CodeLlama. 
Qwen2.5-Coder also produces the most valid tests and achieves the highest success rate, with 7,247 valid tests at 46.03\%, while DeepSeek-Coder generates 5,443 valid tests at 36.67\%, and CodeLlama generates 3,961 valid tests at 34.95\%.
These results indicate that stronger models are better at following instructions to generate valid tests and achieve higher success rates, thereby yielding greater coverage. 
At the same time, we observe that CodeLlama benefits the most from JUnitGenie (34.60\% branch coverage increase and 38.40\% line coverage increase). 
This suggests that while Qwen2.5-Coder has the strongest baseline performance, weaker models such as CodeLlama gain proportionally more from the structured context and distilled knowledge introduced by \tool{}. 
Hence, \tool{} not only improves strong models but also substantially boosts weaker ones, narrowing the performance gap.

Interestingly, Qwen2.5-Coder performs on par with GPT-4o-Mini. 
Under basic prompting, it slightly exceeds GPT-4o-Mini in average coverage.
Under our setting, it achieves the highest branch coverage in four out of 10 projects and the highest line coverage in three. 
These results show that with proper enhancement, strong open-source models can match or even surpass closed-source models, offering a cost-effective solution for large-scale automated test generation.

\find{\textbf{Answers to RQ3:}
\tool{} consistently improves branch and line coverage across all evaluated models by effectively leveraging structured context and refinement.
}

%% file: Tables/Comparison_With_Baselines.tex
\begin{table*}[htbp]
    \scriptsize
    \centering
    \caption{\small Branch/line coverage across approaches. ``–'' denotes no test cases were generated by the tool.}
    \vspace{-0.5mm}
    \label{tab_comparison_with_baseline}
    \resizebox{0.9\textwidth}{!}{
    \begin{tabular}{l*{12}{r}}  
        \toprule
        \multirow{2}{*}{\textbf{Project}} 
        & \multicolumn{2}{c}{\textbf{EvoSuite}} 
        & \multicolumn{2}{c}{\textbf{Randoop}} 
        & \multicolumn{2}{c}{\textbf{ChatTester}} 
        & \multicolumn{2}{c}{\textbf{CoverUp}}
        & \multicolumn{2}{c}{\textbf{HITS}}
        & \multicolumn{2}{c}{\textbf{JUnitGenie}} \\
        \cmidrule(lr){2-3} \cmidrule(lr){4-5} \cmidrule(lr){6-7} \cmidrule(lr){8-9}  \cmidrule(lr){10-11} \cmidrule(lr){12-13}
        & \textbf{Branch} & \textbf{Line}
        & \textbf{Branch} & \textbf{Line}
        & \textbf{Branch} & \textbf{Line}
        & \textbf{Branch} & \textbf{Line}
        & \textbf{Branch} & \textbf{Line}
        & \textbf{Branch} & \textbf{Line} \\
        \midrule
        \rowcolor{gray!35} CodeC         & 64.67\% & 69.90\% & 21.27\% & 28.25\% & 59.65\% & 69.30\% & 66.98\% & 74.93\% & \textbf{75.27\%} & 81.79\% & 72.55\% & \textbf{83.95\%} \\
        \rowcolor{gray!20} Collections   & 51.94\% & 56.92\% & 3.89\%  & 6.35\%  & 54.54\% & 60.36\% & 22.93\% & 23.69\% & 67.70\% & 73.06\% & \textbf{68.89\%} & \textbf{75.58\%} \\
        \rowcolor{gray!35} Compress      & 22.08\% & 19.58\% & 5.45\%  & 6.16\%  & 29.77\% & 29.62\% & 44.72\% & 46.81\% & 45.49\% & 47.34\% & \textbf{54.58\%} & \textbf{57.38\%} \\
        \rowcolor{gray!20} CSV           & 4.62\%  & 7.95\%  & 15.18\% & 22.73\% & 20.79\% & 29.92\% & 66.34\% & 74.62\% & -       & -       & \textbf{76.57\%} & \textbf{85.98\%} \\
        \rowcolor{gray!35} JCore         & 37.01\% & 38.97\% & 6.42\%  & 9.15\%  & 14.86\% & 15.94\% & 21.69\% & 21.83\% & 19.67\% & 22.17\% & \textbf{41.51\%} & \textbf{43.25\%} \\
        \rowcolor{gray!20} JDataBind     & 3.29\%  & 2.39\%  & 3.46\%  & 3.76\%  & 0.48\%  & 0.66\%  & 1.37\%  & 1.72\%  & 1.49\%  & 1.77\%  & \textbf{39.25\%} & \textbf{44.10\%} \\
        \rowcolor{gray!35} JXML          & 1.44\%  & 0.81\%  & 5.44\%  & 7.20\%  & 19.84\% & 25.56\% & \textbf{31.84\%} & \textbf{39.51\%} & 28.32\% & 34.65\% & 27.20\% & 34.92\% \\
        \rowcolor{gray!20} JxPath        & \textbf{49.47\%} & 52.10\% & 6.67\%  & 9.45\%  & 11.43\% & 14.41\% & 27.81\% & 32.46\% & 32.23\% & 37.00\% & 46.96\% & \textbf{53.93\%} \\
        \rowcolor{gray!35} Lang          & 77.73\% & 76.93\% & 7.38\%  & 12.18\% & 35.51\% & 35.65\% & 33.06\% & 36.20\% & 26.77\% & 28.81\% & \textbf{87.10\%} & \textbf{92.41\%} \\
        \rowcolor{gray!20} JodaTime      & 70.52\% & 72.93\% & 27.00\% & 36.21\% & -       & -       & -       & -       & -       & -       & \textbf{74.60\%} & \textbf{79.82\%} \\
        \midrule
        \textbf{Average} & 40.84\% & 41.37\% & 8.86\%  & 12.20\% & 23.70\% & 27.23\% & 30.78\% & 34.78\% & 32.13\% & 36.45\% & \textbf{56.86\%} & \textbf{61.45\%} \\
        \bottomrule 
    \end{tabular}
    }
    \vspace{-5mm}
\end{table*}

%% file: Tables/Ablation_Study.tex
\begin{table}[t]
    \small
    \centering
    \renewcommand{\arraystretch}{1.2} 
    \caption{\small Effectiveness of knowledge acquisition and utilization in LLM-based test generation (CKD: context knowledge distillation, CKE: context knowledge extraction, TCR: test case refinement).}
    \label{tab_ablation_study}
    \vspace{-0.5mm}
    \resizebox{0.5\textwidth}{!}{%
    \begin{tabular}{l*{8}{r}}
        \toprule
        \multirow{2}{*}{\textbf{Project}} 
        & \multicolumn{2}{c}{\textbf{\makecell[l]{\makecell[l]{\ \ \ \ \ CKD\\ w/o  CKE\\ \ \ \ \ \ TCR}}}}
        & \multicolumn{2}{c}{\textbf{\makecell[l]{\makecell[l]{w/o CKD\\ \ \ \ \ \ CKE}}}}
        & \multicolumn{2}{c}{\textbf{w/o CKD}} 
        & \multicolumn{2}{c}{\textbf{\tool{}}} \\
        \cmidrule(lr){2-3} \cmidrule(lr){4-5} \cmidrule(lr){6-7} \cmidrule(lr){8-9}
        & \textbf{Branch} & \textbf{Line}
        & \textbf{Branch} & \textbf{Line}
        & \textbf{Branch} & \textbf{Line}
        & \textbf{Branch} & \textbf{Line} \\
        \midrule
        \rowcolor{gray!35} CodeC         & 41.17\% & 51.26\% & 54.48\% & 68.26\% & 54.48\% & 64.61\% & \textbf{72.55\%} & \textbf{83.95\%} \\
        \rowcolor{gray!20} Collections   & 33.30\% & 41.89\% & 55.53\% & 62.21\% & 53.74\% & 61.15\% & \textbf{68.89\%} & \textbf{75.58\%} \\
        \rowcolor{gray!35} Compress      & 10.48\% & 11.29\% & 29.91\% & 36.43\% & 29.63\% & 31.03\% & \textbf{54.58\%} & \textbf{57.38\%} \\
        \rowcolor{gray!20} CSV           & 8.91\%  & 14.39\% & 50.17\% & 63.64\% & 53.47\% & 72.35\% & \textbf{76.57\%} & \textbf{85.98\%} \\
        \rowcolor{gray!35} JCore         & 5.81\%  & 6.82\%  & 13.55\% & 16.48\% & 15.17\% & 18.85\% & \textbf{41.51\%} & \textbf{43.25\%} \\
        \rowcolor{gray!20} JDataBind     & 14.87\% & 18.12\% & 29.39\% & 34.60\% & 20.37\% & 21.08\% & \textbf{39.25\%} & \textbf{44.10\%} \\
        \rowcolor{gray!35} JXML          & 5.92\%  & 9.36\%  & 8.48\%  & 11.88\% & 15.52\% & 20.16\% & \textbf{27.20\%} & \textbf{34.92\%} \\
        \rowcolor{gray!20} JxPath        & 9.71\%  & 13.28\% & 17.50\% & 24.60\% & 12.62\% & 20.16\% & \textbf{46.96\%} & \textbf{53.93\%} \\
        \rowcolor{gray!35} Lang          & 66.50\% & 68.35\% & 75.29\% & 79.18\% & 65.08\% & 68.65\% & \textbf{87.10\%} & \textbf{92.41\%} \\
        \rowcolor{gray!20} JodaTime      & 49.56\% & 57.34\% & 63.06\% & 71.58\% & 57.37\% & 64.24\% & \textbf{74.60\%} & \textbf{79.82\%} \\
        \midrule
        \textbf{Average} & 24.74\% & 27.84\% & 37.65\% & 42.97\% & 34.44\% & 38.70\% & \textbf{56.86\%} & \textbf{61.45\%} \\
        \bottomrule
    \end{tabular}
    }
\end{table}

%% file: Tables/Different_Foundation_Models.tex
\begin{table*}[t]
    \centering
    \caption{\small Comparison of branch and line coverage using basic prompting (Basic) vs. \tool{} (Ours) across four LLMs.}
    \label{tab_different_foundation_model_reordered}
    \vspace{-1mm}
    \resizebox{0.96\textwidth}{!}{%
\begin{tabular}{
l
!{\vrule width 0.5pt}*{2}{r}
!{\vrule width 0.5pt}*{2}{r}
!{\vrule width 0.5pt}*{2}{r}
!{\vrule width 0.5pt}*{2}{r}
!{\vrule width 1pt}
*{2}{r}
!{\vrule width 0.5pt}*{2}{r}
!{\vrule width 0.5pt}*{2}{r}
!{\vrule width 0.5pt}*{2}{r}
}
        \toprule
        \multirow{2}{*}{\textbf{Project}} 
        & \multicolumn{8}{c!{\vrule width 1pt}}{\textbf{Branch Coverage}}  
        & \multicolumn{8}{c}{\textbf{Line Coverage}} \\
        & \multicolumn{2}{c}{GPT-4o-Mini} 
        & \multicolumn{2}{c}{CodeLlama} 
        & \multicolumn{2}{c}{DeepSeek-Coder} 
        & \multicolumn{2}{c!{\vrule width 1pt}}{Qwen2.5-Coder}  
        & \multicolumn{2}{c}{GPT-4o-Mini} 
        & \multicolumn{2}{c}{CodeLlama} 
        & \multicolumn{2}{c}{DeepSeek-Coder} 
        & \multicolumn{2}{c}{Qwen2.5-Coder} \\
        
        \cmidrule(lr){2-3} \cmidrule(lr){4-5} \cmidrule(lr){6-7} \cmidrule(lr){8-9}
        \cmidrule(lr){10-11} \cmidrule(lr){12-13} \cmidrule(lr){14-15} \cmidrule(lr){16-17}
        & Basic & Ours & Basic & Ours & Basic & Ours & Basic & Ours 
        & Basic & Ours & Basic & Ours & Basic & Ours & Basic & Ours \\
        
        \arrayrulecolor{black}
        \specialrule{1pt}{0pt}{0pt}
        \rowcolor{gray!35} CodeC        & 41.17\% & 72.55\% & 13.82\% & 63.96\% & 28.26\% & 72.36\% & 47.55\% & 79.08\% & 51.26\% & 83.95\% & 25.80\% & 77.99\% & 38.68\% & 79.03\% & 58.11\% & 90.03\% \\
        \rowcolor{gray!20} Collections  & 33.30\% & 68.89\% & 12.66\% & 53.24\% & 15.59\% & 60.52\% & 42.97\% & 68.10\% & 41.89\% & 75.58\% & 15.95\% & 58.97\% & 21.91\% & 71.08\% & 48.97\% & 73.59\% \\
        \rowcolor{gray!35} Compress     & 10.48\% & 54.58\% & 3.91\% & 38.16\% & 7.69\% & 41.51\% & 11.81\% & 54.23\% & 11.29\% & 57.38\% & 5.48\% & 42.02\% & 9.35\% & 45.25\% & 12.09\% & 55.36\% \\
        \rowcolor{gray!20} CSV          & 8.91\% & 76.57\% & 1.98\% & 67.33\% & 16.50\% & 61.06\% & 42.90\% & 76.90\% & 14.39\% & 85.98\% & 1.14\% & 79.17\% & 17.80\% & 75.00\% & 39.02\% & 85.61\% \\
        \rowcolor{gray!35} JCore        & 5.81\% & 41.51\% & 2.38\% & 25.08\% & 4.55\% & 27.86\% & 8.19\% & 38.12\% & 6.82\% & 43.25\% & 2.64\% & 30.43\% & 5.00\% & 31.63\% & 10.38\% & 41.78\% \\
        \rowcolor{gray!20} JDataBind    & 14.87\% & 39.25\% & 4.42\% & 33.09\% & 6.87\% & 34.35\% & 21.57\% & 35.90\% & 18.12\% & 44.10\% & 4.99\% & 38.14\% & 7.91\% & 38.93\% & 23.91\% & 40.61\% \\
        \rowcolor{gray!35} JXML         & 5.92\% & 27.20\% & 1.92\% & 26.56\% & 3.04\% & 24.16\% & 10.08\% & 29.92\% & 9.36\% & 34.92\% & 2.97\% & 33.93\% & 2.43\% & 31.05\% & 13.05\% & 36.90\% \\
        \rowcolor{gray!20} JxPath       & 9.71\% & 46.96\% & 2.38\% & 30.12\% & 8.72\% & 37.19\% & 14.66\% & 46.04\% & 13.28\% & 53.93\% & 3.13\% & 38.31\% & 13.56\% & 43.12\% & 19.78\% & 52.29\% \\
        \rowcolor{gray!35} Lang         & 66.50\% & 87.10\% & 34.21\% & 75.67\% & 47.30\% & 84.08\% & 66.30\% & 87.42\% & 68.35\% & 92.41\% & 37.44\% & 82.78\% & 52.72\% & 87.47\% & 68.05\% & 92.01\% \\
        \rowcolor{gray!20} JodaTime     & 49.56\% & 74.60\% & 21.23\% & 67.50\% & 27.98\% & 73.09\% & 57.10\% & 76.47\% & 57.34\% & 79.82\% & 25.56\% & 72.70\% & 34.59\% & 77.75\% & 60.99\% & 80.94\% \\
        \midrule
        \textbf{Average} & 24.74\% & \textbf{56.86\%} & 10.31\% & \textbf{44.91\%} & 16.20\% & \textbf{49.31\%} & 29.83\% & \textbf{56.38\%} 
                         & 27.84\% & \textbf{61.45\%} & 12.27\% & \textbf{50.67\%} & 19.04\% & \textbf{53.92\%} & 32.03\% & \textbf{60.68\%} \\
        \bottomrule
    \end{tabular}
    }
\vspace{-4mm}
\end{table*}

%% file: 5_Discussion.tex
\section{Discussion}
\input{Tables/Bug_Detection}
\subsection{Potential to Reveal Real-World Bugs} \label{sec_real_world_bugs} 
\tool{} is primarily designed to generate unit tests that achieve high path coverage within focal methods. 
While it is not explicitly intended for triggering bugs, its ability to explore deeper and more diverse control-flow paths increases the chance of exposing edge cases and latent issues.
To assess this potential, we execute all valid tests generated by \tool{} across the 10 projects.
For each failed test execution, we manually examine the expected and actual behaviors to determine if the discrepancy reflects a genuine bug.

We discovered eight issues and reported them to the developers.  
Development teams have reacted to all these issues and four of them are confirmed as real bugs and have been fixed.
We also noted that several of the confirmed bugs were fixed shortly after our reports were submitted, suggesting that test failures and accompanying details provided by \tool{} are useful in helping developers identify and resolve real bugs efficiently.
In the future, we plan to incorporate issue-related knowledge into \tool{} to guide context extraction and distillation, enabling the generation of test cases that can uncover real-world bugs or vulnerabilities.

\subsection{Threats to Validity}
The main limitation of \tool{} is scalability on highly branching control flow.
Path coverage is NP-hard: the number of CFG paths grows exponentially with branches (e.g., $n$ branches yield up to $2^n$ paths)~\cite{dorf2018computers}.
Our analysis shows 99\% of methods in our dataset have fewer than 100 CFG paths, while under 1\% suffer severe path explosion.
Although rare, such methods remain challenging for automated test generation.

Second, some generated tests invoke version-incompatible APIs or unavailable libraries, causing compilation or runtime errors.
While \textit{Feedback Loop Refinement} repairs many cases using diagnostics, some remain invalid since error messages cannot resolve version mismatches or missing dependencies.
In future work, we plan to incorporate environment-specific constraints and pre-installed library knowledge into prompting to better match the runtime environment.

Third, in our evaluation (Section~\ref{sec_evaluation}), we consider tests \emph{valid} if they reach the focal method as they contribute to coverage regardless of whether exceptions indicate bugs or incorrect inputs. 
A later exception may indicate a real bug or an incorrect oracle, risking false positives. 
Nevertheless, our path-sensitive context enables the LLM to infer useful oracles, demonstrated by discovering unknown real-world bugs.
Automatic oracle generation remains an open challenge~\cite{Barr2015TheOP}.
Future work will incorporate oracle-checking techniques into JUnitGenie to produce more accurate oracles.

Finally, results may be influenced by randomness, as EvoSuite and Randoop use randomized search while LLM-based methods rely on probabilistic generation.
We mitigate this by fixing the dataset and environment and checking consistency under four different time budgets. 
Across all budgets, \tool{} consistently achieves the highest branch and line coverage, demonstrating robustness to stochasticity.


%% file: Tables/Bug_Detection.tex


\begin{table}[t]
    \centering
    \caption{\small Bug detection results.}
    \label{tab_bugs_detection}
    \resizebox{0.49\textwidth}{!}{
    \begin{tabular}{lcccccc}
        \toprule
        \textbf{Project} & \textbf{\# Star} & \textbf{Issue ID(s)} & \textbf{Confirm?} & \textbf{Fix Commit} & \textbf{Severity} \\
        \midrule
        CodeC & 469 & \href{https://issues.apache.org/jira/browse/CODEC-328}{CODEC-328} & Yes & \href{https://github.com/apache/commons-codec/commit/da16f363adf7c60d45002602afa3c466ddff427c}{da16f36} & Major \\
        CodeC & 469 & \href{https://issues.apache.org/jira/browse/CODEC-330}{CODEC-330} & Yes & \href{https://github.com/apache/commons-codec/commit/c4d8365e0fccb62dfe6a33a3054247f43be11b83}{c4d8365} & Major \\ 
        CodeC & 469 & \href{https://issues.apache.org/jira/browse/CODEC-331}{CODEC-331} & Yes & \href{https://github.com/apache/commons-codec/commit/59929c391777e60157ae9533f624f8383c007e85}{59929c3} & Major \\
        Lang & 2.8k & \href{https://issues.apache.org/jira/browse/LANG-1771}{LANG-1771} & Yes & \href{https://github.com/apache/commons-lang/commit/5a15ccb62eb88ffa41e6b62ac944066fb121afe8}{5a15ccb} & Major \\
        \bottomrule
    \end{tabular}
    }
    \vspace{-1mm}
\end{table}




%% file: 6_Related_Work.tex
\section{Related Work}
Unit testing~\cite{Olan2003UnitTT, Runeson2006ASO, Zhu1997SoftwareUT,1990StandardGO, sun2022mining} validates whether a program unit (e.g., a method) behaves as expected~\cite{Olan2003UnitTT, Runeson2006ASO, Zhu1997SoftwareUT,1990StandardGO}.
To find errors early in the development process, researchers focus on automatically generating unit tests that cover all branches and lines~\cite{Daka2014ASO, Tassey2002TheEI, Wang2024HITSHL, Klammer2015WritingUT}.
To this end, many automated test generation techniques have been proposed~\cite{Blasi2022CallMM, Kurian2022AutomaticallyGT, Lukasczyk2022PynguinAU, Lukasczyk2021AnES, Scalabrino2018OCELOTAS}, including search-based~\cite{Blasi2022CallMM, DelgadoPrez2023InterEvoTRIE, Harman2010ATA, Harman2015AchievementsOP, McMinn2011SearchBasedST, fraser2011evosuite, Fraser2011EvoSuiteAT, Fraser2013EvoSuiteOT, Fraser20151600FI}, constraint-based~\cite{Csallner2008DySy, Ernst2007TheDS, Xiao2013CharacteristicSO}, symbolic execution-based~\cite{Baldoni2016ASO, CadarUsenixA8, Cha2012UnleashingMO, Chipounov2012TheSP}, and random-based strategies~\cite{Pacheco2007FeedbackDirectedRT, fuzzingbook}.
For example, Fraser et al.~\cite{fraser2011evosuite} proposed EvoSuite, which uses evolutionary algorithms to evolve test cases, while Xiao et al.~\cite{Xiao2013CharacteristicSO}  applied symbolic execution to systematically explore paths.
However, these heuristic methods struggle with complex type constraints and semantic dependencies, lacking the deep reasoning needed to handle inter-procedural logic, and thus often yield limited coverage on complex methods~\cite{Herlim2021EmpiricalSO, McMinn2011SearchBasedST, Tang2023ChatGPTVS, CadarUsenixA8, Xie2009FitnessguidedPE}.

To address these limitations, recent research has explored deep learning (DL) for automated test generation.
One approach trains task-specific DL models on large corpora of human-written tests~\cite{Hu2023IdentifyAU, Rao2023CATLMTL, Alagarsamy2023A3TestAA, Watson2020OnLM, Tufano2020UnitTC, Dinella2021TOGAAN, Nie2023LearningDS, tufano2020unit}, allowing them to learn semantic patterns from code–test pairs.
For example, ATLAS~\cite{Watson2020OnLM} uses neural machine translation to generate assertions, while AthenaTest~\cite{Tufano2020UnitTC} fine-tunes BART~\cite{Chipman2008BARTBA} to generate tests from method context.
However, these methods require large datasets and costly training.
Recent works instead exploit the general-purpose capabilities of LLMs~\cite{Yuan2024EvaluatingAI, Chen2023ChatUniTestAF, Wang2024HITSHL, Ryan2024CodeAwarePA, Jiang2024TowardsUT, Hayet2025ChatAssertLT, Oudraogo2024LLMsAP, Yang2024AdvancingCC, Nan2025TestIG, Gu2025LLMTG, Yin2025EnhancingLA, han2024chase, pan2024large, zhang2025deployability}, generating tests by prompting with the focal method and its context.
For instance, ChatTester~\cite{Yuan2024EvaluatingAI} includes class and field/method signatures, while HITS~\cite{Wang2024HITSHL} adds referenced class declarations.
However, focal methods often depend on external elements such as variables and nested calls~\cite{Yin2025EnhancingLA, Ryan2024CodeAwarePA}.
Without this context, models must infer missing semantics, often yielding invalid or unexecutable tests.
SymPrompt~\cite{Ryan2024CodeAwarePA} partially mitigates this by adding type and dependency information in Python prompts, but lacks structured guidance.
In contrast, \tool{} extracts rich context and distills it into explicit, path-sensitive guidance, enabling executable tests with high coverage across complex paths.
Recently, exLong \cite{Zhang2024exLongGE} also combines LLMs with path-condition analysis for exceptional-behavior tests.
It derives guard predicates from execution traces and fine-tunes a model to synthesize tests that hit \emph{throw} sites.
In contrast, \tool{} targets general path coverage by extracting CFG path–sensitive context and distilling it into structured prompts without fine-tuning.

%% file: 7_Conclusion.tex
\section{Conclusion}

In this work, we present \tool{}, a prototype for path-sensitive test generation in Java.
It extracts contextual code knowledge and distills it into structured instructions to guide LLMs in generating high-coverage tests with minimal cognitive load.
Extensive experiments demonstrate that:
(1) \tool{} outperforms both heuristic and LLM-based baselines, achieving an average of 56.86\% branch coverage and 61.45\% line coverage;
(2) providing abundant context alone is insufficient; effective test generation requires distilling context into explicit instructions;
(3) \tool{} generalizes well across closed- and open-source foundation models;
(4) \tool{} can uncover real-world bugs, highlighting its practical utility.

%% file: 0_Main.bbl
\begin{thebibliography}{10}

\bibitem{netguru2025}
Netguru.
\newblock Is java still used in 2025?
\newblock {\em Netguru Blog}, 2025.
\newblock Accessed: 2025-05-01.

\bibitem{sun2023lazycow}
Xiaoyu Sun, Xiao Chen, Yonghui Liu, John Grundy, and Li~Li.
\newblock Lazycow: A lightweight crowdsourced testing tool for taming android fragmentation.
\newblock In {\em Proceedings of the 31st ACM Joint European Software Engineering Conference and Symposium on the Foundations of Software Engineering}, pages 2127--2131, 2023.

\bibitem{zhao2022autopatch}
Yanjie Zhao, Pei Liu, Xiaoyu Sun, Yue Liu, Yonghui Liu, John Grundy, and Li~Li.
\newblock Autopatch: Learning to generate patches for automatically fixing compatibility issues in android apps.
\newblock {\em Available at SSRN 4254659}, 2022.

\bibitem{hu2023detecting}
Haonan Hu, Yue Liu, Yanjie Zhao, Yonghui Liu, Xiaoyu Sun, Chakkrit Tantithamthavorn, and Li~Li.
\newblock Detecting temporal inconsistency in biased datasets for android malware detection.
\newblock In {\em 2023 38th IEEE/ACM International Conference on Automated Software Engineering Workshops (ASEW)}, pages 17--23. IEEE, 2023.

\bibitem{zhang2025fixing}
Lyuye Zhang, Jiahui Wu, Chengwei Liu, Kaixuan Li, Xiaoyu Sun, Lida Zhao, Chong Wang, and Yang Liu.
\newblock Fixing outside the box: Uncovering tactics for open-source security issue management.
\newblock {\em Proceedings of the ACM on Software Engineering}, 2(ISSTA):2273--2295, 2025.

\bibitem{sun2023taming}
Xiaoyu Sun, Xiao Chen, Yonghui Liu, John Grundy, and Li~Li.
\newblock Taming android fragmentation through lightweight crowdsourced testing.
\newblock {\em IEEE Transactions on Software Engineering}, 49(6):3599--3615, 2023.

\bibitem{Daka2014ASO}
Ermira Daka and Gordon Fraser.
\newblock A survey on unit testing practices and problems.
\newblock {\em 2014 IEEE 25th International Symposium on Software Reliability Engineering}, pages 201--211, 2014.

\bibitem{Klammer2015WritingUT}
Claus Klammer and Albin Kern.
\newblock Writing unit tests: It's now or never!
\newblock {\em 2015 IEEE Eighth International Conference on Software Testing, Verification and Validation Workshops (ICSTW)}, pages 1--4, 2015.

\bibitem{daka2014survey}
Ermira Daka and Gordon Fraser.
\newblock A survey on unit testing practices and problems.
\newblock In {\em 2014 IEEE 25th International Symposium on Software Reliability Engineering}, pages 201--211. IEEE, 2014.

\bibitem{fraser2011evosuite}
Gordon Fraser and Andrea Arcuri.
\newblock Evosuite: automatic test suite generation for object-oriented software.
\newblock In {\em Proceedings of the 19th ACM SIGSOFT symposium and the 13th European conference on Foundations of software engineering}, pages 416--419, 2011.

\bibitem{DelgadoPrez2023InterEvoTRIE}
Pedro Delgado-P{\'e}rez, Aurora Ram{\'i}rez, Kevin~J. Valle-G{\'o}mez, Inmaculada Medina-Bulo, and Jos{\'e}~Ra{\'u}l Romero.
\newblock Interevo-tr: Interactive evolutionary test generation with readability assessment.
\newblock {\em IEEE Transactions on Software Engineering}, 49:2580--2596, 2023.

\bibitem{Harman2010ATA}
Mark Harman and Phil McMinn.
\newblock A theoretical and empirical study of search-based testing: Local, global, and hybrid search.
\newblock {\em IEEE Transactions on Software Engineering}, 36:226--247, 2010.

\bibitem{Csallner2008DySy}
Christoph Csallner, Nikolai Tillmann, and Yannis Smaragdakis.
\newblock Dysy.
\newblock {\em 2008 ACM/IEEE 30th International Conference on Software Engineering}, pages 281--290, 2008.

\bibitem{Ernst2007TheDS}
Michael~D. Ernst, Jeff~H. Perkins, Philip~J. Guo, Stephen McCamant, Carlos Pacheco, Matthew~S. Tschantz, and Chen Xiao.
\newblock The daikon system for dynamic detection of likely invariants.
\newblock {\em Sci. Comput. Program.}, 69:35--45, 2007.

\bibitem{Xiao2013CharacteristicSO}
Xusheng Xiao, Sihan Li, Tao Xie, and Nikolai Tillmann.
\newblock Characteristic studies of loop problems for structural test generation via symbolic execution.
\newblock {\em 2013 28th IEEE/ACM International Conference on Automated Software Engineering (ASE)}, pages 246--256, 2013.

\bibitem{Pacheco2007FeedbackDirectedRT}
Carlos Pacheco, Shuvendu~K. Lahiri, Michael~D. Ernst, and Thomas Ball.
\newblock Feedback-directed random test generation.
\newblock {\em 29th International Conference on Software Engineering (ICSE'07)}, pages 75--84, 2007.

\bibitem{fuzzingbook}
Andreas Zeller, Rahul Gopinath, Marcel Böhme, Gordon Fraser, and Christian Holler.
\newblock {\em The Fuzzing Book}.
\newblock 2019.

\bibitem{Herlim2021EmpiricalSO}
Robert~Sebastian Herlim, Shin Hong, Yunho Kim, and Moonzoo Kim.
\newblock Empirical study of effectiveness of evosuite on the sbst 2020 tool competition benchmark.
\newblock In {\em International Symposium on Search Based Software Engineering}, 2021.

\bibitem{Yuan2024EvaluatingAI}
Zhiqiang Yuan, Mingwei Liu, Shiji Ding, Kaixin Wang, Yixuan Chen, Xin Peng, and Yiling Lou.
\newblock Evaluating and improving chatgpt for unit test generation.
\newblock {\em Proc. ACM Softw. Eng.}, 1:1703--1726, 2024.

\bibitem{Chen2023ChatUniTestAF}
Yinghao Chen, Zehao Hu, Chen Zhi, Junxiao Han, Shuiguang Deng, and Jianwei Yin.
\newblock Chatunitest: A framework for llm-based test generation.
\newblock In {\em SIGSOFT FSE Companion}.

\bibitem{Liao2023A3CodGenAR}
Dianshu Liao, Shidong Pan, Xiaoyu Sun, Xiaoxue Ren, Qing Huang, Zhenchang Xing, Huan Jin, and Qinying Li.
\newblock $\mathbf{A^{3}}$a3-codgen: A repository-level code generation framework for code reuse with local-aware, global-aware, and third-party-library-aware.
\newblock {\em IEEE Transactions on Software Engineering}, 50(12):3369--3384, 2024.

\bibitem{Shi2023LargeLM}
Freda Shi, Xinyun Chen, Kanishka Misra, Nathan Scales, David Dohan, Ed~H. Chi, Nathanael Scharli, and Denny Zhou.
\newblock Large language models can be easily distracted by irrelevant context.
\newblock In {\em International Conference on Machine Learning}, 2023.

\bibitem{Fraser2011EvoSuiteAT}
Gordon Fraser and Andrea Arcuri.
\newblock Evosuite: automatic test suite generation for object-oriented software.
\newblock In {\em ESEC/FSE '11}, 2011.

\bibitem{Pacheco2007RandoopFR}
Carlos Pacheco and Michael~D. Ernst.
\newblock Randoop: feedback-directed random testing for java.
\newblock In {\em Conference on Object-Oriented Programming Systems, Languages, and Applications}, 2007.

\bibitem{Pizzorno2024CoverUpEH}
Juan~Altmayer Pizzorno and E.~Berger.
\newblock Coverup: Effective high coverage test generation for python.
\newblock 2024.

\bibitem{Wang2024HITSHL}
Zejun Wang, Kaibo Liu, Ge~Li, and Zhi Jin.
\newblock Hits: High-coverage llm-based unit test generation via method slicing.
\newblock {\em 2024 39th IEEE/ACM International Conference on Automated Software Engineering (ASE)}, pages 1258--1268, 2024.

\bibitem{Gu2025LLMTG}
Sijia Gu, Noor Nashid, and Ali Mesbah.
\newblock Llm test generation via iterative hybrid program analysis.
\newblock {\em ArXiv}, abs/2503.13580, 2025.

\bibitem{javaparser}
Javaparser.
\newblock \url{https://github.com/javaparser/javaparser}, 2019.

\bibitem{Karakaya2024SootUpAR}
Kadiray Karakaya, Stefan Schott, Jonas Klauke, Eric Bodden, Markus Schmidt, Linghui Luo, and Dongjie He.
\newblock Sootup: A redesign of the soot static analysis framework.
\newblock In {\em International Conference on Tools and Algorithms for Construction and Analysis of Systems}, 2024.

\bibitem{lhotak2003scaling}
Ond{\v{r}}ej Lhot{\'a}k and Laurie Hendren.
\newblock Scaling java points-to analysis using spark.
\newblock In {\em International conference on compiler construction}, pages 153--169. Springer, 2003.

\bibitem{Neo4J}
Neo4j: Build what others can't for.
\newblock \url{https://neo4j.com/}.
\newblock Accessed: 2025-05-30.

\bibitem{sun2021taming}
Xiaoyu Sun, Li~Li, Tegawend{\'e}~F Bissyand{\'e}, Jacques Klein, Damien Octeau, and John Grundy.
\newblock Taming reflection: An essential step toward whole-program analysis of android apps.
\newblock {\em ACM Transactions on Software Engineering and Methodology (TOSEM)}, 30(3):1--36, 2021.

\bibitem{Wei2022ChainOT}
Jason Wei, Xuezhi Wang, Dale Schuurmans, Maarten Bosma, Ed~H. Chi, F.~Xia, Quoc Le, and Denny Zhou.
\newblock Chain of thought prompting elicits reasoning in large language models.
\newblock {\em ArXiv}, abs/2201.11903, 2022.

\bibitem{our_repo}
Artifact: Source code, prompts, datasets, and experimental results for this paper \href{https://github.com/Dianshu-Liao/JUnitGenie}{https://github.com/Dianshu-Liao/JUnitGenie}.

\bibitem{javac}
javac - java programming language compiler.
\newblock \url{https://docs.oracle.com/javase/7/docs/technotes/tools/solaris/javac.html}.

\bibitem{Just2014Defects4JAD}
Ren{\'e} Just, Darioush Jalali, and Michael~D. Ernst.
\newblock Defects4j: a database of existing faults to enable controlled testing studies for java programs.
\newblock In {\em International Symposium on Software Testing and Analysis}, 2014.

\bibitem{Mak2024AutomaticBR}
Ching~Hang Mak and Shing~Chi Cheung.
\newblock Automatic build repair for test cases using incompatible java versions.
\newblock {\em Inf. Softw. Technol.}, 172:107473, 2024.

\bibitem{Watkinson2024ComparingAA}
Myles Watkinson and Alexander E.~I. Brownlee.
\newblock Comparing apples and oranges? investigating the consistency of cpu and memory profiler results across multiple java versions.
\newblock {\em Automated Software Engineering}, 31:1--34, 2024.

\bibitem{Schott2024JavaBN}
Stefan Schott, Serena~Elisa Ponta, Wolfram Fischer, Jonas Klauke, and Eric Bodden.
\newblock Java bytecode normalization for code similarity analysis (artifact).
\newblock {\em Dagstuhl Artifacts Ser.}, 10:20:1--20:3, 2024.

\bibitem{GPT4omini}
OpenAI.
\newblock Gpt-4o-mini.
\newblock \url{https://openai.com/index/gpt-4o-mini-advancing-cost-efficient-intelligence/}, 2024.

\bibitem{Rozire2023CodeLO}
Baptiste Rozi{\`e}re, Jonas Gehring, Fabian Gloeckle, Sten Sootla, Itai Gat, Xiaoqing Tan, Yossi Adi, Jingyu Liu, Tal Remez, J{\'e}r{\'e}my Rapin, Artyom Kozhevnikov, I.~Evtimov, Joanna Bitton, Manish~P Bhatt, Cris tian Cant{\'o}n~Ferrer, Aaron Grattafiori, Wenhan Xiong, Alexandre D'efossez, Jade Copet, Faisal Azhar, Hugo Touvron, Louis Martin, Nicolas Usunier, Thomas Scialom, and Gabriel Synnaeve.
\newblock Code llama: Open foundation models for code.
\newblock {\em ArXiv}, abs/2308.12950, 2023.

\bibitem{Guo2024DeepSeekCoderWT}
Daya Guo, Qihao Zhu, Dejian Yang, Zhenda Xie, Kai Dong, Wentao Zhang, Guanting Chen, Xiao Bi, Yu~Wu, Y.~K. Li, Fuli Luo, Yingfei Xiong, and Wenfeng Liang.
\newblock Deepseek-coder: When the large language model meets programming - the rise of code intelligence.
\newblock {\em ArXiv}, abs/2401.14196, 2024.

\bibitem{Hui2024Qwen25CoderTR}
Binyuan Hui, Jian Yang, Zeyu Cui, Jiaxi Yang, Dayiheng Liu, Lei Zhang, Tianyu Liu, Jiajun Zhang, Bowen Yu, Kai Dang, An~Yang, Rui Men, Fei Huang, Shanghaoran Quan, Xingzhang Ren, Xuancheng Ren, Jingren Zhou, and Junyang Lin.
\newblock Qwen2.5-coder technical report.
\newblock {\em ArXiv}, abs/2409.12186, 2024.

\bibitem{dorf2018computers}
Richard~C Dorf.
\newblock {\em Computers, software engineering, and digital devices}.
\newblock CRC Press, 2018.

\bibitem{Barr2015TheOP}
Earl~T. Barr, Mark Harman, Phil McMinn, Muzammil Shahbaz, and Shin Yoo.
\newblock The oracle problem in software testing: A survey.
\newblock {\em IEEE Transactions on Software Engineering}, 41:507--525, 2015.

\bibitem{Olan2003UnitTT}
Michael Olan.
\newblock Unit testing: test early, test often.
\newblock {\em J. Comput. Sci. Coll.}, 19:319--328, 2003.

\bibitem{Runeson2006ASO}
Per Runeson.
\newblock A survey of unit testing practices.
\newblock {\em IEEE Software}, 23:22--29, 2006.

\bibitem{Zhu1997SoftwareUT}
Hong Zhu, Patrick A.~V. Hall, and John H.~R. May.
\newblock Software unit test coverage and adequacy.
\newblock {\em ACM Computing Surveys (CSUR)}, 29:366 -- 427, 1997.

\bibitem{1990StandardGO}
Standard glossary of software engineering terminology.
\newblock 1990.

\bibitem{sun2022mining}
Xiaoyu Sun, Xiao Chen, Yanjie Zhao, Pei Liu, John Grundy, and Li~Li.
\newblock Mining android api usage to generate unit test cases for pinpointing compatibility issues.
\newblock In {\em Proceedings of the 37th IEEE/ACM International Conference on Automated Software Engineering}, pages 1--13, 2022.

\bibitem{Tassey2002TheEI}
Gregory Tassey.
\newblock The economic impacts of inadequate infrastructure for software testing.
\newblock 2002.

\bibitem{Blasi2022CallMM}
Arianna Blasi, Alessandra Gorla, Michael~D. Ernst, and Mauro Pezz{\`e}.
\newblock Call me maybe: Using nlp to automatically generate unit test cases respecting temporal constraints.
\newblock {\em Proceedings of the 37th IEEE/ACM International Conference on Automated Software Engineering}, 2022.

\bibitem{Kurian2022AutomaticallyGT}
Elson Kurian, Daniela Briola, Pietro Braione, and Giovanni Denaro.
\newblock Automatically generating test cases for safety-critical software via symbolic execution.
\newblock {\em J. Syst. Softw.}, 199:111629, 2022.

\bibitem{Lukasczyk2022PynguinAU}
Stephan Lukasczyk and Gordon Fraser.
\newblock Pynguin: Automated unit test generation for python.
\newblock {\em 2022 IEEE/ACM 44th International Conference on Software Engineering: Companion Proceedings (ICSE-Companion)}, pages 168--172, 2022.

\bibitem{Lukasczyk2021AnES}
Stephan Lukasczyk, Florian Kroi{\ss}, and Gordon Fraser.
\newblock An empirical study of automated unit test generation for python.
\newblock {\em Empirical Software Engineering}, 28:1--46, 2021.

\bibitem{Scalabrino2018OCELOTAS}
Simone Scalabrino, Giovanni Grano, Dario~Di Nucci, Michele Guerra, Andrea~De Lucia, Harald~C. Gall, and Rocco Oliveto.
\newblock Ocelot: A search-based test-data generation tool for c.
\newblock {\em 2018 33rd IEEE/ACM International Conference on Automated Software Engineering (ASE)}, pages 868--871, 2018.

\bibitem{Harman2015AchievementsOP}
Mark Harman, Yue Jia, and Yuanyuan Zhang.
\newblock Achievements, open problems and challenges for search based software testing.
\newblock {\em 2015 IEEE 8th International Conference on Software Testing, Verification and Validation (ICST)}, pages 1--12, 2015.

\bibitem{McMinn2011SearchBasedST}
Phil McMinn.
\newblock Search-based software testing: Past, present and future.
\newblock {\em 2011 IEEE Fourth International Conference on Software Testing, Verification and Validation Workshops}, pages 153--163, 2011.

\bibitem{Fraser2013EvoSuiteOT}
Gordon Fraser and Andrea Arcuri.
\newblock Evosuite: On the challenges of test case generation in the real world.
\newblock {\em 2013 IEEE Sixth International Conference on Software Testing, Verification and Validation}, pages 362--369, 2013.

\bibitem{Fraser20151600FI}
Gordon Fraser and Andrea Arcuri.
\newblock 1600 faults in 100 projects: automatically finding faults while achieving high coverage with evosuite.
\newblock {\em Empirical Software Engineering}, 20:611--639, 2015.

\bibitem{Baldoni2016ASO}
Roberto Baldoni, Emilio Coppa, Daniele~Cono D'Elia, Camil Demetrescu, and Irene Finocchi.
\newblock A survey of symbolic execution techniques.
\newblock {\em ACM Computing Surveys (CSUR)}, 51:1 -- 39, 2016.

\bibitem{CadarUsenixA8}
Cristian Cadar, Daniel Dunbar, and Dawson~R. Engler.
\newblock Usenix association 8th usenix symposium on operating systems design and implementation 209 klee: Unassisted and automatic generation of high-coverage tests for complex systems programs.

\bibitem{Cha2012UnleashingMO}
Sang~Kil Cha, Thanassis Avgerinos, Alexandre Rebert, and David Brumley.
\newblock Unleashing mayhem on binary code.
\newblock {\em 2012 IEEE Symposium on Security and Privacy}, pages 380--394, 2012.

\bibitem{Chipounov2012TheSP}
Vitaly Chipounov, Volodymyr Kuznetsov, and George Candea.
\newblock The s2e platform: Design, implementation, and applications.
\newblock {\em ACM Trans. Comput. Syst.}, 30:2:1--2:49, 2012.

\bibitem{Tang2023ChatGPTVS}
Yutian Tang, Zhijie Liu, Zhichao Zhou, and Xiapu Luo.
\newblock Chatgpt vs sbst: A comparative assessment of unit test suite generation.
\newblock {\em IEEE Transactions on Software Engineering}, 50:1340--1359, 2023.

\bibitem{Xie2009FitnessguidedPE}
Tao Xie, Nikolai Tillmann, Jonathan de~Halleux, and Wolfram Schulte.
\newblock Fitness-guided path exploration in dynamic symbolic execution.
\newblock {\em 2009 IEEE/IFIP International Conference on Dependable Systems \& Networks}, pages 359--368, 2009.

\bibitem{Hu2023IdentifyAU}
Identify and update test cases when production code changes: A transformer-based approach.
\newblock {\em 2023 38th IEEE/ACM International Conference on Automated Software Engineering (ASE)}, pages 1111--1122, 2023.

\bibitem{Rao2023CATLMTL}
Nikitha Rao, Kush Jain, Uri Alon, Claire~Le Goues, and Vincent~J. Hellendoorn.
\newblock Cat-lm training language models on aligned code and tests.
\newblock {\em 2023 38th IEEE/ACM International Conference on Automated Software Engineering (ASE)}, pages 409--420, 2023.

\bibitem{Alagarsamy2023A3TestAA}
Saranya Alagarsamy, Chakkrit~Kla Tantithamthavorn, and Aldeida Aleti.
\newblock A3test: Assertion-augmented automated test case generation.
\newblock {\em ArXiv}, abs/2302.10352, 2023.

\bibitem{Watson2020OnLM}
Cody Watson, Michele Tufano, Kevin Moran, Gabriele Bavota, and Denys Poshyvanyk.
\newblock On learning meaningful assert statements for unit test cases.
\newblock {\em 2020 IEEE/ACM 42nd International Conference on Software Engineering (ICSE)}, pages 1398--1409, 2020.

\bibitem{Tufano2020UnitTC}
Michele Tufano, Dawn Drain, Alexey Svyatkovskiy, Shao~Kun Deng, and Neel Sundaresan.
\newblock Unit test case generation with transformers.
\newblock {\em ArXiv}, abs/2009.05617, 2020.

\bibitem{Dinella2021TOGAAN}
Elizabeth Dinella, Gabriel Ryan, and Todd Mytkowicz.
\newblock Toga: A neural method for test oracle generation.
\newblock {\em 2022 IEEE/ACM 44th International Conference on Software Engineering (ICSE)}, pages 2130--2141, 2021.

\bibitem{Nie2023LearningDS}
Pengyu Nie, Rahul Banerjee, Junyi~Jessy Li, Raymond~J. Mooney, and Milo{\v{s}} Gligori{\'c}.
\newblock Learning deep semantics for test completion.
\newblock {\em 2023 IEEE/ACM 45th International Conference on Software Engineering (ICSE)}, pages 2111--2123, 2023.

\bibitem{tufano2020unit}
Michele Tufano, Dawn Drain, Alexey Svyatkovskiy, Shao~Kun Deng, and Neel Sundaresan.
\newblock Unit test case generation with transformers and focal context.
\newblock {\em arXiv preprint arXiv:2009.05617}, 2020.

\bibitem{Chipman2008BARTBA}
Hugh~A. Chipman, Edward~I. George, and Robert~E. McCulloch.
\newblock Bart: Bayesian additive regression trees.
\newblock {\em The Annals of Applied Statistics}, 4:266--298, 2008.

\bibitem{Ryan2024CodeAwarePA}
Gabriel Ryan, Siddhartha Jain, Mingyue Shang, Shiqi Wang, Xiaofei Ma, Murali~Krishna Ramanathan, and Baishakhi Ray.
\newblock Code-aware prompting: A study of coverage guided test generation in regression setting using llm.
\newblock {\em Proc. ACM Softw. Eng.}, 1:951--971, 2024.

\bibitem{Jiang2024TowardsUT}
Zongze Jiang, Ming Wen, Jialun Cao, Xuanhua Shi, and Hai Jin.
\newblock Towards understanding the effectiveness of large language models on directed test input generation.
\newblock {\em 2024 39th IEEE/ACM International Conference on Automated Software Engineering (ASE)}, pages 1408--1420, 2024.

\bibitem{Hayet2025ChatAssertLT}
Ishrak Hayet, Adam Scott, and Marcelo d’Amorim.
\newblock Chatassert: Llm-based test oracle generation with external tools assistance.
\newblock {\em IEEE Transactions on Software Engineering}, 51:305--319, 2025.

\bibitem{Oudraogo2024LLMsAP}
Wendk{\^u}uni~C. Ou{\'e}draogo, Kader Kabore, Haoye Tian, Yewei Song, Anil Koyuncu, Jacques Klein, David Lo, and T{\'e}gawend{\'e}~F. Bissyand{\'e}.
\newblock Llms and prompting for unit test generation: A large-scale evaluation.
\newblock {\em 2024 39th IEEE/ACM International Conference on Automated Software Engineering (ASE)}, pages 2464--2465, 2024.

\bibitem{Yang2024AdvancingCC}
Chen Yang, Junjie Chen, Bin Lin, Ziqi Wang, and Jianyi Zhou.
\newblock Advancing code coverage: Incorporating program analysis with large language models.
\newblock {\em Transactions on Software Engineering and Methodology}, 2024.

\bibitem{Nan2025TestIG}
Zifan Nan, Zhaoqiang Guo, Kui Liu, and Xin Xia.
\newblock Test intention guided llm-based unit test generation.
\newblock {\em 2025 IEEE/ACM 47th International Conference on Software Engineering (ICSE)}, pages 1026--1038, 2025.

\bibitem{Yin2025EnhancingLA}
Xin Yin, Chao Ni, Xinrui Li, Liushan Chen, Guojun Ma, and Xiaohu Yang.
\newblock Enhancing llm's ability to generate more repository-aware unit tests through precise contextual information injection.
\newblock {\em ArXiv}, abs/2501.07425, 2025.

\bibitem{han2024chase}
Linyi Han, Shidong Pan, Zhenchang Xing, Jiamou Sun, Sofonias Yitagesu, Xiaowang Zhang, and Zhiyong Feng.
\newblock Do chase your tail! missing key aspects augmentation in textual vulnerability descriptions of long-tail software through feature inference.
\newblock {\em IEEE Transactions on Software Engineering}, 2024.

\bibitem{pan2024large}
Shidong Pan, Tianchen Guo, Lihong Zhang, Pei Liu, Zhenchang Xing, and Xiaoyu Sun.
\newblock A large-scale investigation of semantically incompatible apis behind compatibility issues in android apps.
\newblock {\em arXiv preprint arXiv:2406.17431}, 2024.

\bibitem{zhang2025deployability}
Tianyi Zhang, Shidong Pan, Zejun Zhang, Zhenchang Xing, and Xiaoyu Sun.
\newblock Deployability-centric infrastructure-as-code generation: An llm-based iterative framework.
\newblock {\em arXiv preprint arXiv:2506.05623}, 2025.

\bibitem{Zhang2024exLongGE}
Jiyang Zhang, Yu~Liu, Pengyu Nie, Junyi~Jessy Li, and Milo{\v{s}} Gligori{\'c}.
\newblock exlong: Generating exceptional behavior tests with large language models.
\newblock {\em 2025 IEEE/ACM 47th International Conference on Software Engineering (ICSE)}, pages 1462--1474, 2024.

\end{thebibliography}
